# Dormitory of Physical and Engineering Sciences:

## *Sleeping Beauties May Be Sleeping Innovations*

### *Part 1: Basic Properties, Cognitive Environment, Characteristics of the Princes*

Anthony F. J. van Raan
Centre for Science and Technology Studies, Leiden University
Wassenaarseweg 52, P.O. Box 9555
2300 RB Leiden, The Netherlands
vanraan@cwts.leidenuniv.nl

### *Abstract*

*A 'Sleeping Beauty in Science' is a publication that goes unnoticed ('sleeps') for a long time and then, almost suddenly, attracts a lot of attention ('is awakened by a prince'). In this paper we investigate important properties of Sleeping Beauties, particularly to find out to what extent Sleeping Beauties are application-oriented and thus are potential Sleeping Innovations. In this study we focus primarily on physics (including materials science and astrophysics) and present first results for chemistry and for engineering & computer science. We find that more than half of the SBs are application-oriented. Therefore, it is important to investigate the reasons for and processes related to delayed recognition. First we analyze basic properties of the SBs such as the time-dependent distribution, author characteristics (names of authors, country, institution), as well as the journals and fields of the SBs are analyzed. Next we develop a new approach in which the cognitive environment of the SBs is analyzed, based on the mapping of Sleeping Beauties using their citation links and conceptual relations, particularly co-citation mapping. In this way we investigate the research themes in which the SBs are 'used' and possible causes of why the premature work in the SBs becomes topical, i.e., the trigger of the awakening of the SBs. This approach is tested with a blue skies SB and an application-oriented SB. We think that the mapping procedures discussed in this paper are not only important for bibliometric analyses. They also provide researchers with useful, interactive tools to discover both relevant older work as well as new developments, for instance in themes related to Sleeping Beauties that are also Sleeping Innovations.*

## Introduction

A 'Sleeping Beauty in Science' is a publication that goes unnoticed ('sleeps') for a long time and then, almost suddenly, attracts a lot of attention ('is awakened by a prince'). Our earlier study (van Raan 2004) focused primarily on the occurrence of Sleeping Beauties, and particularly the derivation of an 'awakening probability function' from the empirical data and the identification of the 'most extreme' Sleeping Beauty. The following characteristics of Sleeping Beauties were found: (1) the probability of awakening after a deep declines rapidly with a power law for longer sleeping periods; (2) for a less deep sleep, the length of the sleeping period matters less for the probability of awakening; and (3) the probability for higher awakening intensities decreases extremely rapidly with an steep power law independent of both length as well as depth of sleep. The above findings mean that there is in fact a continuous range from 'mild' to 'extreme' Sleeping Beauties.

Pioneering work on Sleeping Beauties (SBs) using the concept of delayed recognition was done by Eugene Garfield (Garfield 1970, 1989, 1990). Stent (1972) discusses 'prematurity' in scientific discovery. Recent studies focus on the probability of becoming highly cited in later years (Glänzel, Schlemmer, Thijs 2003); the occurrence of delayed recognition (Glänzel, Garfield 2004); SBs in psychology (Lange 2005) and in physics ('revived classics', Redner 2005; an example of awaked application-oriented work, Marx 2014); the probability of SBs (Burrell 2005); a typology of behavioral patterns in citation histories of SBs (Braun, Glänzel, Schubert 2010; Lachance, Larivière 2014); extreme historical cases (van Calster 2012); the reasons for awakening (Wang, Ma, Chen, Rao 2012); SBs with a short leaping immediately after publication (Li, Ye 2012; Li 2014); different types of SBs in a specific journal (Kozak 2013); SBs and 'durability' of scientific publications in general and its effects on research performance assessments by citation analysis (Costas, van Leeuwen, van Raan 2010, 2011, 2013); SBs in the work of Nobel Prize winners and the dependence of the awake intensity on the citation distribution within the sleeping period (Li, Shi, Zhao, Ye 2014); the identification of the 'princes' (Li, Yu, Zhang, Zhang 2014).

In this paper we present the results of an extensive analysis of Sleeping Beauties in physics, chemistry, and engineering & computer science[1] in order to find out whether in the set of discovered SBs there are application-oriented papers. In such cases we in fact identify delayed possible innovations. In order not to complicate the paper by tripling the amount of tables and figures, we present the underlying data, tables and figures for chemistry and engineering in a separate file 'Additional data en results for Chemistry and Engineering'. We include in this paper the main results for chemistry and engineering in the analysis of recent SBs.

The structure of this paper is as follows. In the next section we will discuss the data collection with a novel search algorithm, the analytical method to define Sleeping Beauties, the definition of the three main fields, and the measuring procedure. Given the complexity of a well-defined identification of SBs, we present a number of examples of how we make a selection of specific sets of SBs. Next we will discuss the results. First an analysis of the basic properties of the SBs such as the time-dependent distribution, author characteristics (names of authors, country, institution), as well as the journals and fields of the SBs. Particularly the latter is an indication of the application-oriented character of an SB. Secondly we will apply advanced citation-network analysis and mapping instruments recently developed in our institute. With these analytical tools we are able to construct and display the cognitive environment of the SBs and to detect direct or indirect connections between SBs and related papers. Also, particularly by the mapping procedure, the research themes of the SBs become visible which provides a further important possibility to identify application-oriented SBs. Finally, we analyze with the same network and mapping methods the citing papers of relatively recent, and after the wakening highly cited physics SBs. In this way we investigate the research communities in which the SBs are 'used' and possible causes of why the premature work in the SBs becomes topical, i.e., the trigger of the awakening of the SBs.

[1] Throughout this paper we will refer to physics, chemistry and engineering & computer science as the three main fields.

## Data, Method and Measuring Procedure

### *Data and Variables to Define Sleeping Beauties*

Following our earlier approach (van Raan 2004) we developed a fast and efficient Sleeping Beauty search algorithm written in SQL which can be applied to the CWTS enhanced Web of Science (WoS) database. With this algorithm we can tune the following four main variables: (1) *length of the sleep* in years after publication; (2) *depth of sleep* in terms of a maximum citation rate during the sleeping period; (3) *awake* period in years after the sleeping period; and (4) *awake intensity* in terms of a minimum citation rate during the awake period. Furthermore, the algorithm allows selection of sets of WoS journal categories and thus restrict the search for Sleeping Beauties to a specific (main) field of science. We stress that although we work with discrete values of the variables, all variables can be tuned through any possible range of values, so that a continuum of Sleeping Beauties is found, ranging from 'mild' to 'extreme' ones. The total period in which the SBs are searched for is 1980-2013[2]. This period implies that we search for SBs in a total set of around 40,600,000 publications. Nevertheless, this long period of almost 35 years still restricts the possibilities of finding SBs: we will only find SBs published from 1980. And it is obvious that, for instance, for SBs with a sleeping period of ten years and an awake period of ten years (so covering in total 20 years), the last publication year for such SBs will be 1994.

Clearly very many combinations of sleeping length, depths of sleep, awake period, and awake intensities can be chosen. This enables the analysis of how the number of SBs depends on these four variables. Such an analysis on the occurrence of SBs was carried out in our first work on SBs (van Raan 2004). In this study the main goal is to discover application-oriented SBs and to improve our understanding of the awakening. We make the following choices for the numerical values of the four variables.

First, we use 2 sleeping periods with length $\boldsymbol{s}$=5 and 10 years, respectively, with publication years starting in 1980. Second, for each of both sleeping periods we take three values for the depth of the sleep: the maximum citation rate (corrected for self-citations) during the sleeping period $\boldsymbol{cs}_{max}$=0 (complete coma), 0.5 (very deep sleep) and 1.0 (deep sleep). Third, for counting the citations during the awake period the search algorithm allows a minimum ($\boldsymbol{a}_{min}$) and a maximum ($\boldsymbol{a}_{max}$) time period. With, for instance, $\boldsymbol{a}_{min}$=2 only those SBs will be found for which citations can be counted in a period of at least 2 years after the defined sleeping period until the end of the total period (2013).

As a consequence however, the measured awake period with the same $\boldsymbol{a}_{min}$ value will be different for SBs with different publication years. For instance, for a SB with publication year 1990, the ten year sleeping period will end in 1999 and with $\boldsymbol{a}_{min}$=2 the awake period runs from 2000 to 2013, which is 13 years. For an SB however with publication year 2000, the ten year sleeping period will end in 2009 and the awake period with $\boldsymbol{a}_{min}$=2 will run from 2010 to 2013, thus only 4 years. SBs with publication year 2003 and a sleeping period of ten years will awake in 2012 but they will not be detected with

[2] The data analysis is carried out with the CWTS bibliometric database which is an improved and enriched version of the WoS database. Publication and citation data are available from 1980. The CWTS bibliometric database allows corrections for self-citations with high precision and provides a highly accurate unification of author names and institutions. We refer to the section Data Infrastructure in http://www.cwts.nl/About-CWTS.

$\boldsymbol{a}_{min}$=2 because only 1 awake year is available for citation counting. By using an $\boldsymbol{a}_{max}$ value we can confine the measured awake period to a specific fixed period. For instance with $\boldsymbol{a}_{min}$=0 and $\boldsymbol{a}_{max}$=2 we count citations in the two years immediately after the sleeping period and for $\boldsymbol{a}_{min}$=2 and $\boldsymbol{a}_{max}$=5 citations are counted for the years 2, 3, 4 and 5 after the sleeping period. Setting $\boldsymbol{a}_{min}$= $\boldsymbol{a}_{max}$ allows counting citations in one specific period of fixed length after the sleeping period. For instance with $\boldsymbol{a}_{min}$=$\boldsymbol{a}_{max}$=2, 5 and 10, citations are counted in a period of 2, 5 and 10 years, respectively, after the sleeping period.

The fourth variable is the 'awake intensity', i.e. the citation rate during the defined awake period. We use the minimum citation rates $\boldsymbol{ca}_{min}$=5, 10, 20, and 50. Thus, in this study the most extreme SBs sleep 10 years ($\boldsymbol{s}$=10) in coma (no citations at all during the sleeping period, $\boldsymbol{cs}_{max}$=0) and has at least on average 50 citations per year ($\boldsymbol{ca}_{min}$=50) during a period of 10 years after the sleep ($\boldsymbol{a}_{min}$=$\boldsymbol{a}_{max}$=10). In contrast, the 'lightest' SBs sleep 5 years ($\boldsymbol{s}$=5) in a deep sleep (maximum 1 citation per year on average, $\boldsymbol{cs}_{max}$=1.0) and has at least on average 5 citations per year ($\boldsymbol{ca}_{min}$=5) during a period of 2 years after the sleep ($\boldsymbol{a}_{min}$=$\boldsymbol{a}_{max}$=2).

### *Measuring Procedure*

In total the above specifications of the four variables lead to 72 observations. On average, the algorithm needs about 90 seconds for each observation. We used a measuring procedure as presented in Table 1. In order to explain the measuring procedure and to get a feeling for the numbers, the results for physics are given in the table. The results for chemistry and engineering & computer science are given in the separate file 'Additional data en results for Chemistry and Engineering', Table A1[3]. We clearly observe that there are many more Sleeping Beauties if we decrease (1) the sleeping time, and/or (2) the depth of the sleep, and/or (3) the awake intensity. We refer to our earlier work in which a mathematical expression is deduced from empirical findings for the number of SBs as a function of the three above variables (van Raan 2004). Because we can change all variables in small steps our measurements allow a search over a broad continuous range.

We take the data in the first gray column as an example. We see that in the entire period (1980-2013) there are in total 989,205 physics publications that have a sleeping period of 5 years ($\boldsymbol{s}$=5) with a 'complete coma', i.e., no citations at all, and thus maximum citation rate $\boldsymbol{cs}_{max}$=0, and for which an awake period of 2 years immediately after the sleeping period ($\boldsymbol{a}_{min}$=$\boldsymbol{a}_{max}$=2) is defined. Thus, a total time span of 7 years anywhere in the entire period 1980-2013, the last possible publication year is 2007. Notice that 989,205 is not the total number of physics publications in the period 1980-2007, which is about 3,500,000, but the total number of publications in this period with citation rate $\boldsymbol{cs}_{max}$=0 during the first five years after publication (sleeping period) and any number of citations in the two years after the sleeping period.

[3] Throughout this paper, we will indicate tables and figures in this separate file 'Additional data en results for Chemistry and Engineering', with an A, for instance Table A1, Figure A1, etc.

<table>
<tr><td>cs(max)=0</td><td></td><td></td><td></td></tr>
<tr><td></td><td></td><td>s = 5</td><td>10</td></tr>
<tr><td>a(min)=a(max)=</td><td>2</td><td>989205|52</td><td>555276|3</td></tr>
<tr><td>ca(min)=5</td><td>5</td><td>862401|52</td><td>473572|4</td></tr>
<tr><td></td><td>10</td><td>653483|71</td><td>330189|3</td></tr>
<tr><td>a(min)=a(max)=</td><td>2</td><td>989205|3</td><td>555276|0</td></tr>
<tr><td>ca(min)=10</td><td>5</td><td>862401|5</td><td>473572|1</td></tr>
<tr><td></td><td>10</td><td>653483|12</td><td>330189|1</td></tr>
<tr><td>a(min)=a(max)=</td><td>2</td><td>989205|0</td><td>555276|0</td></tr>
<tr><td>ca(min)=20</td><td>5</td><td>862401|1</td><td>473572|0</td></tr>
<tr><td></td><td>10</td><td>653483|2</td><td>330189|0</td></tr>
<tr><td>a(min)=a(max)=</td><td>2</td><td>989205|0</td><td>555276|0</td></tr>
<tr><td>ca(min)=50</td><td>5</td><td>862401|0</td><td>473572|0</td></tr>
<tr><td></td><td>10</td><td>653483|1</td><td>330189|0</td></tr>
</table>

<table>
<tr><td>cs(max)=0.5</td><td></td><td></td><td></td></tr>
<tr><td></td><td></td><td>s = 5</td><td>10</td></tr>
<tr><td>a(min)=a(max)=</td><td>2</td><td>1836346|664</td><td>1468682|182</td></tr>
<tr><td>ca(min)=5</td><td>5</td><td>1578310|466</td><td>1233577|122</td></tr>
<tr><td></td><td>10</td><td>1181271|453</td><td>845821|125</td></tr>
<tr><td>a(min)=a(max)=</td><td>2</td><td>1836346|21</td><td>1468682|5</td></tr>
<tr><td>ca(min)=10</td><td>5</td><td>1578310|32</td><td>1233577|10</td></tr>
<tr><td></td><td>10</td><td>1181271|67</td><td>845821|15</td></tr>
<tr><td>a(min)=a(max)=</td><td>2</td><td>1836346|2</td><td>1468682|1</td></tr>
<tr><td>ca(min)=20</td><td>5</td><td>1578310|5</td><td>1233577|3</td></tr>
<tr><td></td><td>10</td><td>1181271|8</td><td>845821|4</td></tr>
<tr><td>a(min)=a(max)=</td><td>2</td><td>1836346|0</td><td>1468682|0</td></tr>
<tr><td>ca(min)=50</td><td>5</td><td>1578310|1</td><td>1233577|0</td></tr>
<tr><td></td><td>10</td><td>1181271|3</td><td>845821|0</td></tr>
</table>

<table>
<tr><td>cs(max)=1.0</td><td></td><td></td><td></td></tr>
<tr><td></td><td></td><td>s = 5</td><td>10</td></tr>
<tr><td>a(min)=a(max)=</td><td>2</td><td>2409619|3994</td><td>1817650|871</td></tr>
<tr><td>ca(min)=5</td><td>5</td><td>2051175|2355</td><td>1518125|488</td></tr>
<tr><td></td><td>10</td><td>1522091|1801</td><td>1035533|389</td></tr>
<tr><td>a(min)=a(max)=</td><td>2</td><td>2409619|87</td><td>1817650|20</td></tr>
<tr><td>ca(min)=10</td><td>5</td><td>2051175|130</td><td>1518125|40</td></tr>
<tr><td></td><td>10</td><td>1522091|204</td><td>1035533|41</td></tr>
<tr><td>a(min)=a(max)=</td><td>2</td><td>2409619|2</td><td>1817650|1</td></tr>
<tr><td>ca(min)=20</td><td>5</td><td>2051175|14</td><td>1518125|7</td></tr>
<tr><td></td><td>10</td><td>1522091|26</td><td>1035533|7</td></tr>
<tr><td>a(max)=a(min)</td><td>2</td><td>2409619|0</td><td>1817650|0</td></tr>
<tr><td>ca(min)=50</td><td>5</td><td>2051175|1</td><td>1518125|0</td></tr>
<tr><td></td><td>10</td><td>1522091|4</td><td>1035533|0</td></tr>
</table>

*Table 1: Results of the measuring procedure with 72 observations for physics.*

Of these 989,205 publications, only 52 have an awake intensity of at least on average 5 citations per year, i.e., minimum citation rate $\boldsymbol{ca}_{\min}$=5 in the two years after the sleeping period. For a sleeping period of 10 years and again $\boldsymbol{cs}_{\max}$=0 (second gray column) we find 555,276 physics publications with $\boldsymbol{a}_{\min}$=$\boldsymbol{a}_{\max}$=2. This number is lower than the number we found with a sleeping period of 5 years because there are less publications for which a total time span of 12 years within the period 1980-2013 is available. We see that there are only 3 publications out of these 555,276 that meet the threshold of awake intensity $\boldsymbol{ca}_{\min}$=5. In a number of cases we observe that the number of SBs increases with increasing measured awake period. This can be explained by considering the situation that in a short awake period after the sleeping period (such as $\boldsymbol{a}_{\min}$=$\boldsymbol{a}_{\max}$=2) publication may not yet have reached their highest impact but they do so in later years.

In this study we focus on a selection of several specific sets of SBs chosen from all the above possible sets. In particular we are interested in the set of SBs with at most a deep sleep ($\boldsymbol{cs}_{\max}$=1.0) during 10 years ($\boldsymbol{s}$=10), an awake period of 10 years ($\boldsymbol{a}_{\min}$=$\boldsymbol{a}_{\max}$=10) during which the SBs have a minimum citation rate $\boldsymbol{ca}_{\min}$=5.0. Thus, 1994 is the last year for publications having a twenty year time span until 2013.

The number of these physics SBs is 389 (out of 1,035,533, see Table 1). This is a sufficiently large set for analysis. Notice that within this set we can find more 'extreme' SBs: the maximum citation rate during sleep is 1.0, so within this set we can find all SBs with a citation rate between 0 and 1.0. And similarly, with the minimum citation rate during the awake period of 5.0, we can find all SBs with a citation rate of for instance 10, 20 or 50, or higher. Only sleeping period and awake period are fixed. Thus, having collected the above set of 389 SBs, all other sets indicated in Table 1 with bold face and larger figures are subsets of the 389 SBs set. Thus, a separate data collection for these subsets is only necessary if we want to know the total number of publications in the given sleeping and awake period as explained above.

Also for chemistry and engineering we identify SBs with the same variables as the 389 physics SBs ($\boldsymbol{s}$=10, $\boldsymbol{cs}_{\max}$=1.0, $\boldsymbol{a}_{\min}$=$\boldsymbol{a}_{\max}$=10, $\boldsymbol{ca}_{\min}$=5.0, in short notion used throughout this paper: [10, 1.0, 10, 5.0]). For chemistry we find 265 of such SBs (out of 1,168,181, see Table A1), and for engineering & computer science 367 (out of 1,155,532, see Table A1), thus numbers of the same order of magnitude.

To illustrate our investigation with a striking example, we show in Fig.1 the citation track of quite an extreme case: a sleeping period of almost 15 years followed by a very sharp increase. Why was this publication (Takeda & Shiraishi 1994) found although our search algorithm was programmed for a sleeping period of 10 years? The reason is that in the period of the following 10 years –which is according to the setting for the search algorithm the awake period- in the last year of this period the number of citations is so high that the average citation rate for the whole 'awake' period is above 10.0.

It is intriguing, given the main purpose of this study, that this Sleeping Beauty is application–oriented, although it is theoretical work and published in the top basic physics journal Physical Review B. The work is about a new development in the study of flat hexagonal ('honeycomb') structures of silicon atoms (silicene) as nanostructures for semiconductors and has recently attracted attention of graphene research community. Another clear indication that this work is application-oriented is the affiliation of the authors: NTT, Nippon Telegraph & Telephone Corporation. In Fig.A1 an example of extreme SBs in chemistry and in engineering & computer science are show shown.

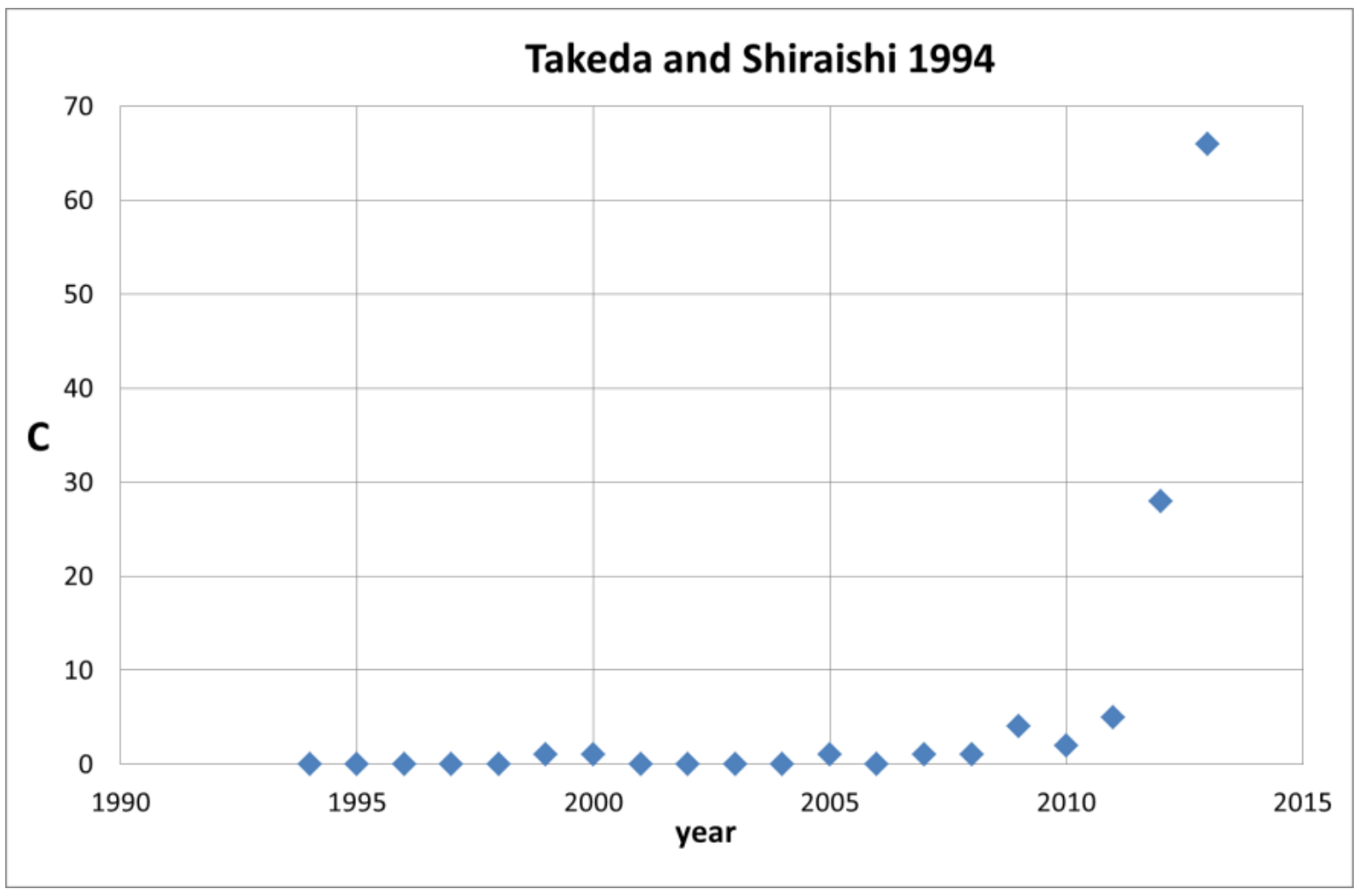

*Figure 1: The citation history of the Takeda & Shiraishi (1994) Sleeping Beauty. The number of citations (self-citations excluded) is indicated with C.*

We also conducted two analyses of relatively recent SBs. The first analysis concerns the SBs in the set of the 389 SBs with publication years 1992, 1993, 1994. In particular, we investigate characteristics of the citing papers ('the princes') of these SBs. The second analysis is focused on the combination of physics, chemistry and engineering & computer science for a shorter sleeping period $\boldsymbol{s}$=5, coma sleep ($\boldsymbol{cs}_{max}$=0), an awake period of 5 years ($\boldsymbol{a}_{min}$=$\boldsymbol{a}_{max}$=5) and an awake intensity $\boldsymbol{ca}_{min}$=5.0, in short notation [5, 0.0, 5, 5.0]. For this set the last year is 2004. In total we found 148 SBs of which 63 in the period 2000-2004. In order not to overload this paper, the latter analysis will be discussed in a follow-up paper (van Raan 2015).

### *Definition of the main fields*

We define physics, chemistry and engineering & computer science as a set of WoS journal categories (throughout this paper indicated as fields), see Table 2 for physics. The definitions for chemistry and engineering & computer science are given in Table A2.

| Physics | |
|---|---|
| **WoS field code and name** | |
| 1 | ACOUSTICS |
| 20 | ASTRONOMY & ASTROPHYSICS |
| 27 | BIOPHYSICS |
| 35 | THERMODYNAMICS |
| 152 | MATERIALS SCIENCE, BIOMATERIALS |
| 153 | MATERIALS SCIENCE, CHARACTERIZATION & TESTING |
| 154 | MATERIALS SCIENCE, COATINGS & FILMS |
| 155 | MATERIALS SCIENCE, COMPOSITES |
| 156 | MATERIALS SCIENCE, TEXTILES |
| 159 | METEOROLOGY & ATMOSPHERIC SCIENCES |
| 168 | NUCLEAR SCIENCE & TECHNOLOGY |
| 175 | OPTICS |
| 185 | PHYSICS, APPLIED |
| 187 | PHYSICS, FLUIDS & PLASMAS |
| 188 | PHYSICS, ATOMIC, MOLECULAR & CHEMICAL |
| 189 | PHYSICS, MULTIDISCIPLINARY |
| 190 | PHYSICS, CONDENSED MATTER |
| 192 | PHYSICS, NUCLEAR |
| 193 | PHYSICS, PARTICLES & FIELDS |
| 195 | PHYSICS, MATHEMATICAL |

*Table 2: Definition of the main field physics based on WoS journal categories.*

Thus, physics, chemistry and engineering & computer science are main fields composed of (sub)fields as indicated in the tables. The WoS journal-category codes are field identifiers used in our CWTS enhanced bibliometric WoS-based data system. The search algorithm selects the Sleeping Beauties in the fields defined by these identifiers. Notice that we use a broad definition of physics by including materials science and astronomy & astrophysics.

## Basic Properties of Sleeping Beauties

### *Number of Sleeping Beauties as a function of time*

One could argue that the probability for a publication to become a SB is less than, say, 20 years ago. Thus, in recent times the number of SBs will be less than 20 years ago. This argument is based on the consideration that nowadays the opportunities for accessibility and promotion of publications are much larger than for instance in the 1980's and as a result less publications will go unnoticed for a longer time.

With our SB search algorithm we investigated the time-dependent distribution of SBs. As discussed in the foregoing section, we selected for all three main fields the SBs with a sleeping period of 10 years, a maximum citation rate of 1.0 during the sleeping period, an awake period of 10 years after the sleep and a minimum citation rate of 10 during the whole awake period. The selection provides us sets of 389 physics SBs, 265 chemistry SBs and 367 engineering & computer science SBs. For these sets we calculated the distribution over the publication years. Contrary to the above argumentation, we see an increasing absolute number of SBs toward more recent years, from around 15 to 20 per year in the early 1980's to around 30 to 45 per year in the early 1990's. For chemistry and engineering & computer science we find similar distributions, although for the latter main field the number for the last years of the observation period are higher, around 40-60.

However, the total number of physics publications in the data-system increased exponentially from around 75,000 in 1980 to 118,000 in 1994 (and 250,000 in 2013). This increase has two effects: the more publications are *published* in a given year, the higher the probability of Sleeping Beauties. But an increasing number of publications also means that there are more publications available as potential *citing* papers, also from other fields, which makes the probability for SBs lower. These effects are not very influential: we found that a simple normalization on the basis of the total number of publications per year does not change the general picture. The results for physics are shown in Fig.2 and for chemistry and engineering & computer science in Fig.A2.

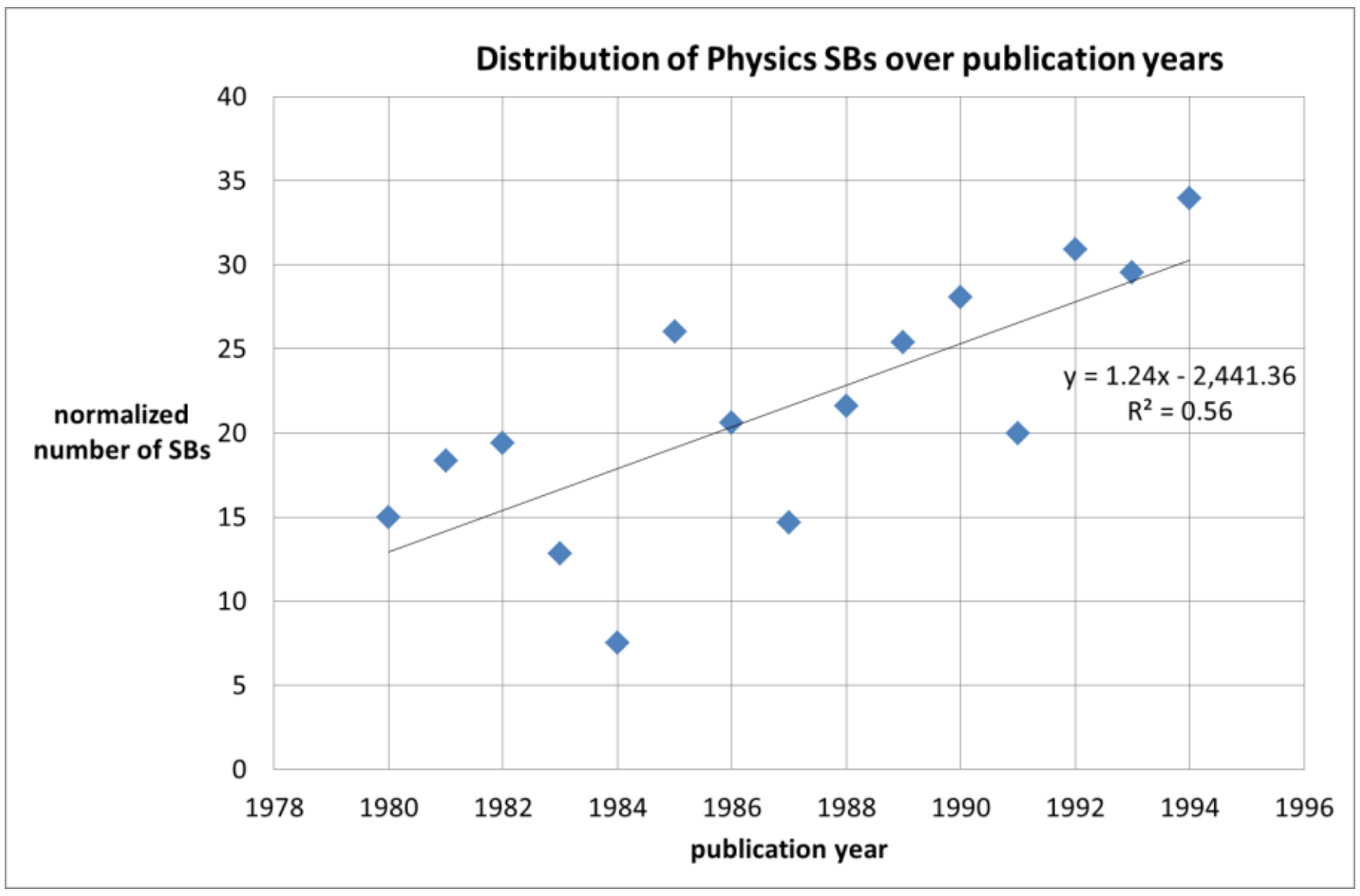

*Figure 2: Normalized number of physics SBs by publication year.*

Although the statistical significance of the trend lines is low, the figures for the three main fields suggest that the increase in absolute numbers of SBs is larger for physics as compared to chemistry but lower than for engineering & computer science.

### *Locations of Sleeping Beauties*

Where do the SBs come from? Are they based on work in developing countries and less known universities, or do countries with a strong science system and top-universities also produce SBs? In order to answer this question we analyzed the locations of the 389 physics, 265 chemistry and 367 engineering & computer science SBs. We present the top-15 countries of the physics SBs distribution in Table 3. The results for all three main fields are given in Table A3, for countries with 2 or more SBs. Given the international collaboration, also countries mentioned as second, third, etc. address are included.

| Country | Number of SBs | % of total |
|---|---|---|
| USA | 147 | 38.0 |
| JAPAN | 48 | 12.4 |
| USSR | 26 | 6.7 |
| UK | 21 | 5.4 |
| FED REP GER | 18 | 4.7 |
| FRANCE | 17 | 4.4 |
| ITALY | 15 | 3.9 |
| CANADA | 14 | 3.6 |
| INDIA | 11 | 2.8 |
| PEOPLES R CHINA | 9 | 2.3 |
| SWEDEN | 8 | 2.1 |
| SPAIN | 8 | 2.1 |
| NETHERLANDS | 8 | 2.1 |
| POLAND | 7 | 1.8 |
| ISRAEL | 7 | 1.8 |

*Table 3: Distribution of physics SBs over countries (the first 15 countries of the distribution).*

Table 3 shows that the distribution of SBs over countries does not deviate much from the general distribution of publications over countries. Next we present the institutions from which the SBs originate, see Table 4 for physics. Institutions with at least 5 SBs are listed. The results for all three main fields (physics, chemistry, engineering & computer science) are shown in Table A4 (institutions with 2 or more SBs).

| Institution | Number of SBs | % of total |
|---|---|---|
| MIT | 11 | 2.8 |
| UNIV CAMBRIDGE | 8 | 2.1 |
| INDIANA UNIV | 7 | 1.8 |
| UNIV TOKYO | 6 | 1.6 |
| UNIV CALIF S BARBARA | 6 | 1.6 |
| IST NAZL FIS NUCL | 6 | 1.6 |
| CALTECH | 6 | 1.6 |
| ACAD SCI USSR | 6 | 1.6 |
| TOKYO INST TECHNOL | 5 | 1.3 |
| SUNY STONY BROOK | 5 | 1.3 |
| PRINCETON UNIV | 5 | 1.3 |
| CITY UNIV NEW YORK | 5 | 1.3 |
| BULGARIAN ACAD SCI | 5 | 1.3 |

*Table 4: Institutions with 5 or more physics SBs.*

As we see in Table 4, top-universities are well represented and it is certainly not the case that physics SBs originate predominantly from less known institutions. For chemistry and engineering & computer science the number of institutions with a high number of SBs is lower than in physics. For physics we do not find any private sector institution with two or more SBs. In chemistry a business company (Upjohn) is on top of the list and in engineering & computer science we find several business companies, with IBM in the top. An explanation of this difference is probably that physics is the most fundamental main field, whereas chemistry and engineering are generally more application-oriented. In the next section we will further investigate this issue on the basis of the research fields to which the SBs belong.

***Are Sleeping Beauties more basic than applied research?***

One is perhaps more inclined to believe that SBs relate to more fundamental, rather exotic, 'blue skies' work (at least for the time the work was performed) and less to application-oriented work. Therefore we analyzed the research fields of the SBs. We present all physics fields with more than 10 SBs in Table 5. The results for all three main fields (physics, chemistry, engineering & computer science) are shown in Table A5, in which all (sub)fields with more than 2 SBs are given. Notice that we find more fields than those fields used for the definition of specific main fields. For instance in Table 5 we see the field electrical & electronic engineering. This field is used for the definition of main field engineering & computer science and not for the definition of main field physics, yet it shows up 18 times as a field in which physics SBs are published. The reason is that journals can be attributed to more than one field. In this particular case, the 18 physics SBs are published in journals which are attributed to both fields such as applied physics or materials science as well as to the field electrical & electronic engineering.

| **Field** | **Number of SBs** | **% of total** |
|---|---|---|
| PHYSICS, MULTIDISCIPLINARY | 85 | 22.0 |
| PHYSICS, APPLIED | 53 | 13.7 |
| PHYSICS, CONDENSED MATTER | 44 | 11.4 |
| PHYSICS, PARTICLES & FIELDS | 42 | 10.9 |
| OPTICS | 33 | 8.5 |
| PHYSICS, ATOMIC, MOLECULAR & CHEMICAL | 29 | 7.5 |
| PHYSICS, MATHEMATICAL | 27 | 7.0 |
| ASTRONOMY & ASTROPHYSICS | 27 | 7.0 |
| METEOROLOGY & ATMOSPHERIC SCIENCES | 19 | 4.9 |
| ENGINEERING, ELECTRICAL & ELECTRONIC | 18 | 4.7 |
| BIOPHYSICS | 18 | 4.7 |
| CHEMISTRY, PHYSICAL | 17 | 4.4 |
| MATERIALS SCIENCE, MULTIDISCIPLINARY | 15 | 3.9 |
| THERMODYNAMICS | 12 | 3.1 |
| MECHANICS | 12 | 3.1 |
| PHYSICS, FLUIDS & PLASMAS | 11 | 2.8 |
| ENVIRONMENTAL SCIENCES | 11 | 2.8 |
| BIOCHEMISTRY & MOLECULAR BIOLOGY | 11 | 2.8 |
| ENGINEERING, BIOMEDICAL | 10 | 2.6 |

*Table 5: Number of SBs in the different physics fields with ten or more SBs.*

In total, the 389 physics SBs are classified 610 times, of which 210 times in applied fields such as applied physics, meteorology & atmospheric sciences, electrical and electronic engineering, biophysics, materials science. Thus, 53% of the 389 physics SBs are classified in at least one applied (sub)field. This means that around half of the Sleeping are also potential Sleeping Innovations!

| Journal | Number of SBs | % of total |
|---|---|---|
| PHYSICS LETTERS B | 26 | 6.7 |
| PHYSICAL REVIEW D | 17 | 4.4 |
| NUCLEAR PHYSICS B | 17 | 4.4 |
| PHYSICAL REVIEW B | 15 | 3.9 |
| PHYSICAL REVIEW LETTERS | 11 | 2.8 |
| JOURNAL OF APPLIED PHYSICS | 11 | 2.8 |
| PHYSICS LETTERS A | 9 | 2.3 |
| PHYSICAL REVIEW A | 8 | 2.1 |
| JOURNAL OF THE ELECTROCHEMICAL SOCIETY | 7 | 1.8 |
| JOURNAL OF PHYSICS A MATHEMATICAL AND GENERAL | 7 | 1.8 |
| JOURNAL OF THE PHYSICAL SOCIETY OF JAPAN | 6 | 1.6 |
| JOURNAL OF CHEMICAL PHYSICS | 6 | 1.6 |
| JETP LETTERS | 6 | 1.6 |
| CHEMICAL PHYSICS LETTERS | 6 | 1.6 |
| APPLIED OPTICS | 6 | 1.6 |
| JAPANESE JOURNAL OF APPLIED PHYSICS PART 1 | 5 | 1.3 |

*Table 6: Distribution of the physics SBs over journals with 5 or more SBs.*

The SBs were published in a large number, around 100, of journals. Table 6 present the journals with more than 5 SBs. These 16 journals cover 42% of the total number of SBs. Several applied research journals are listed such as Journal of Applied Physics, Journal of the Electrochemical Society, Applied Optics, and the Japanese Journal of Applied Physics Part 1. Further analysis of the journals in which the 389 physics SBs were published confirms that around half of these SBs are application-oriented. The lists with journals with two or more SBs for all three main fields (physics, chemistry, and engineering & computer science) are given in Table A6. First results indicate that in chemistry even a larger part (70%) of the SBs is application-oriented, and in engineering & computer science in fact by definition all SBs are application-oriented.

### *Authors of Sleeping Beauties*

An further intriguing question is whether a Sleeping Beauty represents one rare piece of work in the total oeuvre of an author, or that authors produce more than one SB. The latter situation would indicate that at least for the author(s) concerned the SB is not a 'byproduct' but the result of a well-planned research theme. Furthermore there is the related question whether the Sleeping Beauty is the result of a 'lonely wolf' action of just one author, or that SBs are also the result of co-authoring scientists. In Table 7 we present the physics authors who published three or more SBs. Authors who are mentioned together with another author are co-authors. For instance, Kostelecky and Samuel are co-authors in 6 SBs, thus the number of SBs as well as the number of co-authorships is 6. Romans has 5 SBs of which 3 with co-authors, but these co-authors do not have at least 3 publications together with Romans, so they are not included in this table. The results for all three main fields can be found in Table A7 in which authors with more than two SBs are given[4].

[4] For interested readers the author can make the data of this physics SB set available.

| Authors | Number of SBs | Co-authorships |
|---|---|---|
| KOSTELECKY VA, SAMUEL S | 6 | 6 |
| ROMANS LJ | 5 | 3 |
| NOMURA S | 4 | 3 |
| WARD WW | 3 | 2 |
| VIRBHADRA KS | 3 | 0 |
| TOWNSEND PK | 3 | 2 |
| KOSOWSKY A, TURNER MS | 3 | 3 |
| DINE M | 3 | 0 |

*Table 7: Physics authors with at least three SBs.*

In Fig.3 we present the distribution of the number of co-authors per Sleeping Beauty (blue diamonds) as compared with the number of co-authors per physics paper in general. For this letter distribution we took the three journals with the most SBs: Physics Letters B, Physical Review D, and Nuclear Physics B (see Table 6) and created a set of 500 papers in these journals in each of the years 1980, 1985, 1990 and 1994, in order to cover the time span of the 389 SBs. We normalized this distribution to the top of the distribution for the SBs for the ease of comparison.

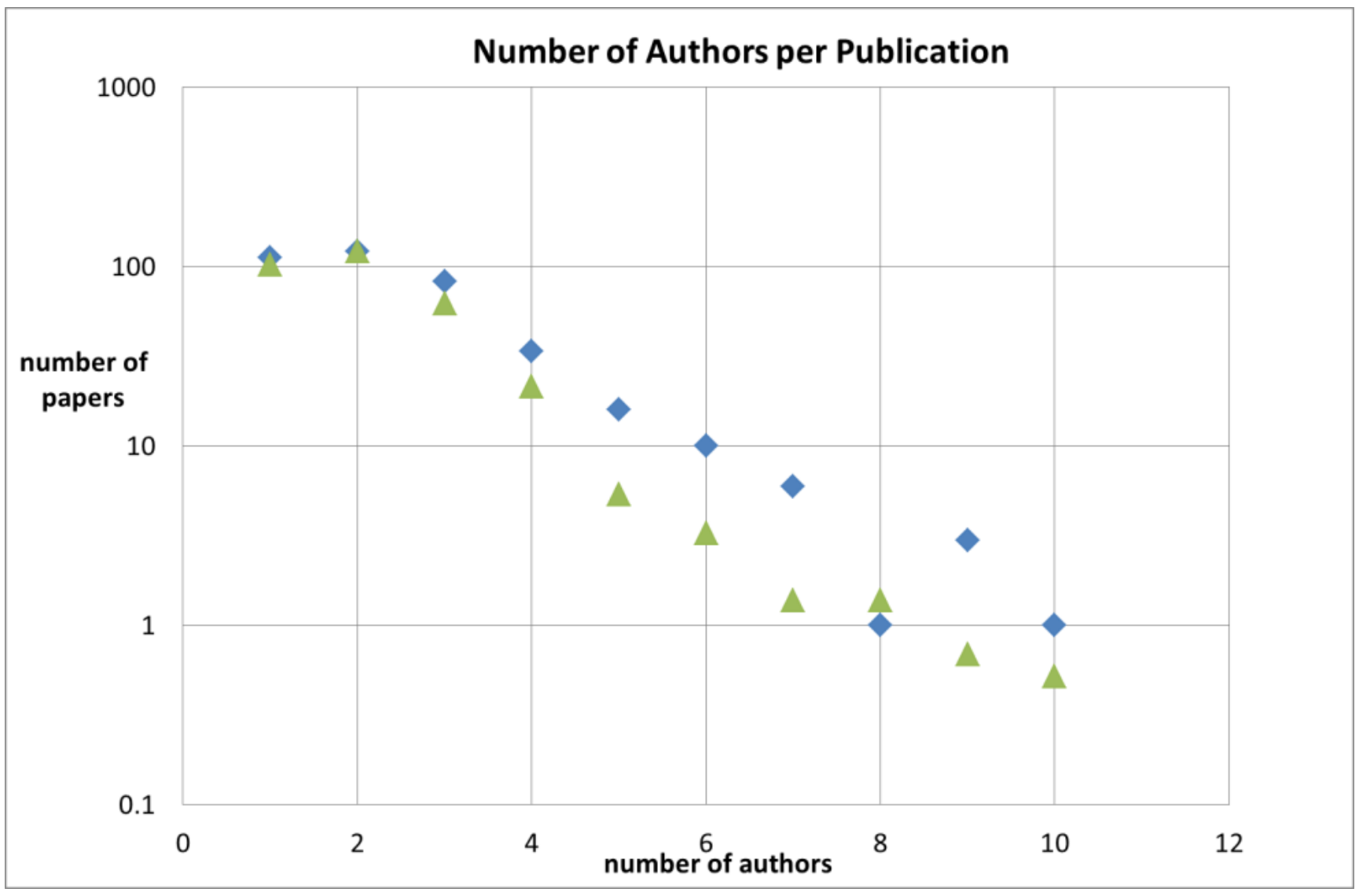

*Figure 3: Distribution of the number of authors per Sleeping Beauty (blue diamonds), and the number of authors per physics paper in general (light green triangles) normalized to the top of the distribution for the SBs.*

Remarkably, the relative number of SBs with up to ten co-authors tends to be somewhat larger as compared to physics papers in general. Nevertheless, the far majority of physics SBs are papers with 1 to 3 authors. A major difference between the SBs distribution and the general distribution is the very long tail in the general distribution (not shown in Fig.3): there are many papers in the three physics journals with 40 of more authors, as can be expected for high energy experimental physics. The highest number of co-authors for the SBs is ten. Evidently, there are no Sleeping Beauties in 'big science', particularly in the many-author high energy experimental physics research. The results for chemistry and for engineering & computer science are given in Fig.A3.

### *Citation histories of Sleeping Beauties*

An important characteristic of a Sleeping Beauty is her path of life after the sleeping period, in terms of citations. From our physics SBs set studied in the foregoing sections (n=389) we selected those SBs with a lower maximum citation rate ($\boldsymbol{cs}_{max}$= 0.5 instead of 1.0) during the 10 years sleep as well as with a higher minimum citation rate ($\boldsymbol{ca}_{min}$=10.0 instead of 5.0) during the ten years awake period. In total we find 15 of these extreme physics SBs. They are a subset (variables [10, 0.5, 10, 10.0]) of the larger set (variables [10, 1.0, 10, 5.0]) of the 389 SB. The distribution of these 15 SBs as a function of time (year of publication) is presented in Fig.4. The figure suggests an increase of the extreme SBs during the more recent years, but given the small number and the rather scattered distribution this is hardly significant. In Fig. A4 we present the results for chemistry and for engineering & computer science.

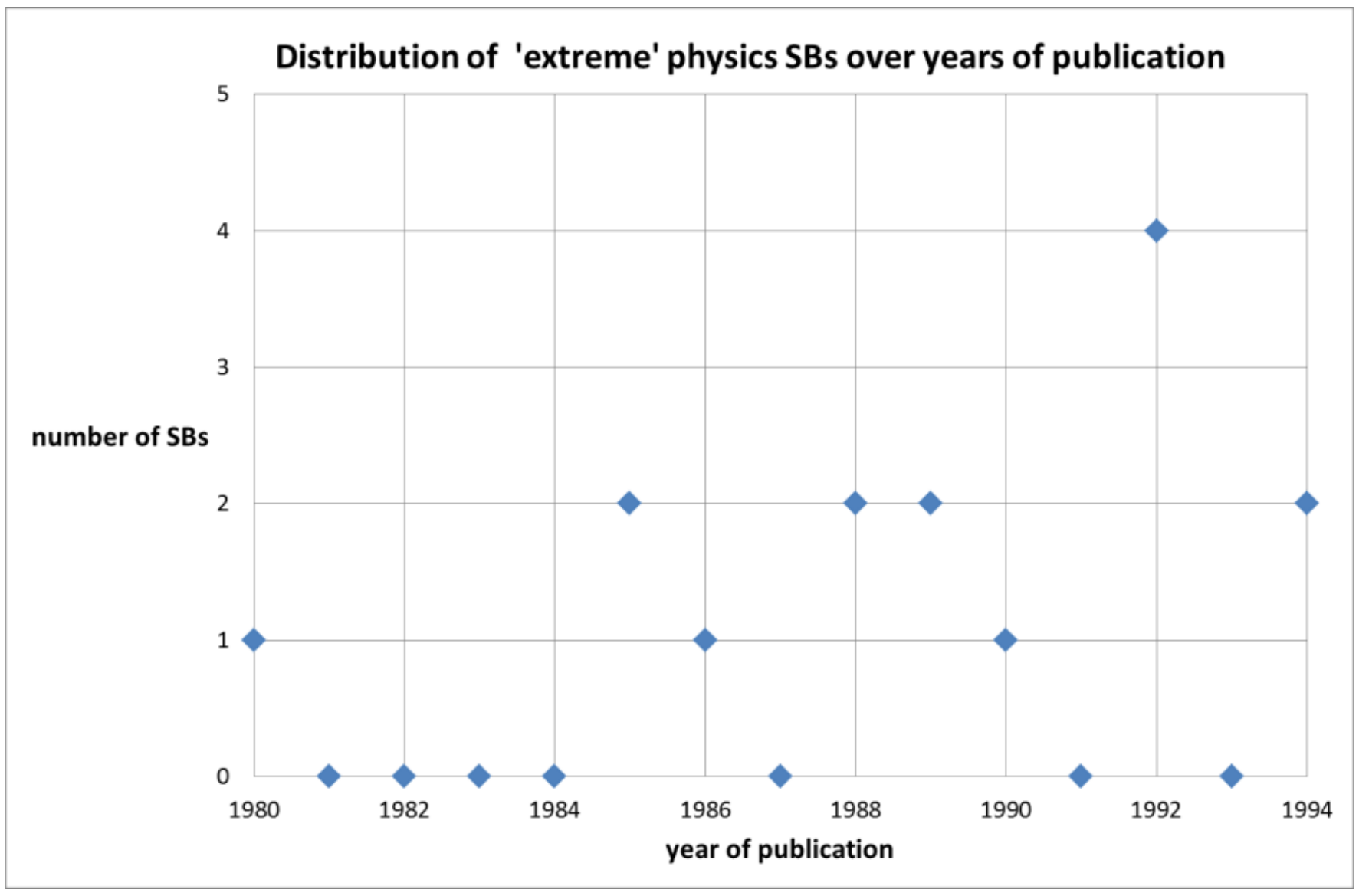

*Figure 4: Number of extreme physics SBs by year of publication.*

We find, at least for these 15 extreme SBs, three main, distinct types of the citation development ('citation history') of these SBs: (1) after the 10 years sleep period a steep rise followed by a fast decrease within 10 years (n=2); (2) after the 10 years sleep period a steep rise followed by a slow decrease within 10 years, (n=8, thus about half of the set); (2) after the sleep period a relatively 'quiet' but continuous increase during the 10 years or even more (n=4). We present an example of each of these types in the Fig.5: the SBs of Simon (1989), Romans (1989b), and Dieks (1988), respectively. We present similar examples for chemistry and for engineering & computer science in Fig.A5.

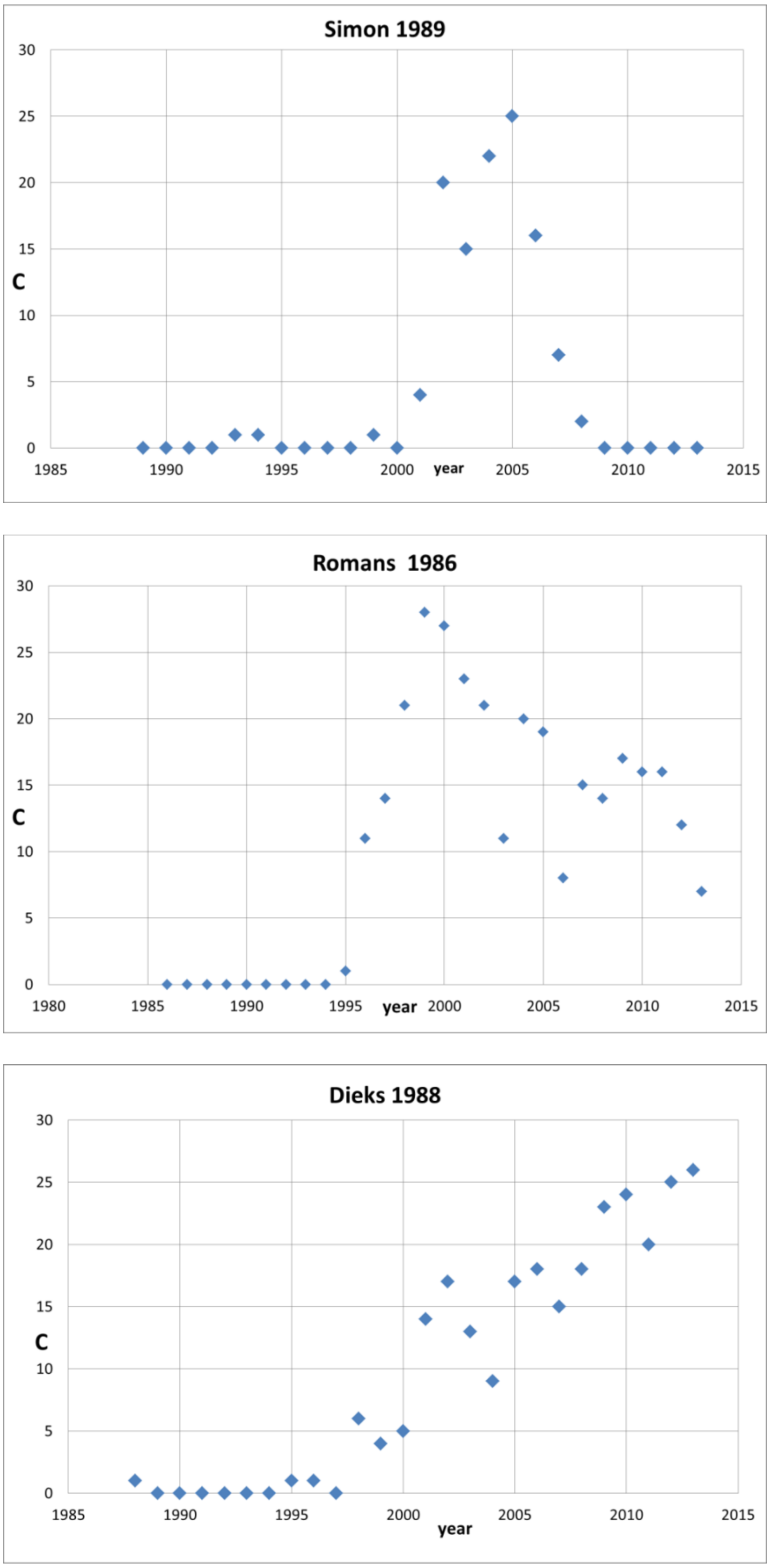

*Figure 5: Examples of the three main types of SB citation histories. Upper part: type a; middle part: type b; lower part: type c. The number of citations (self-citations excluded) is indicated with C.*

## Cognitive Environment of Sleeping Beauties

### *Main lines of the approach*

In the foregoing section we showed that authors may have several SBs so that at least at the level of individual scientists there are connections between different SBs. There are two major bibliometric approaches to investigate the cognitive environment of publications, and here SBs in particular. The first one is citation-based, the second is concept-based.

First the citation-based approach. As any other publication, SBs have links with other (earlier) publications by their references and it is interesting to find out whether there are SBs that have references in common. This might reveal 'families' of SBs, in bibliometric terms these are bibliographically coupled SBs. And the other way around, these common references are co-cited by SBs.

In the second approach we use natural language processing (text mining) to extract the important, publication-specific concepts (terms such as keywords or noun phrases) from the titles and abstracts of a set of SBs. By measuring all co-occurrences of any possible pair of concepts, co-word maps can be created in which the conceptual structure of the research represented by the set of SBs is visualized. For a recent discussion of the concept mapping methodology we refer to Waltman, van Eck and van Raan (2014).

For both approaches we used the recently developed CWTS bibliometric instruments CitNetExplorer and the VOS-viewer. The CitNetExplorer is a software instrument specifically designed for analyzing and visualizing citation networks of scientific literature and it can be uploaded with sets of publication records directly from the Web of Science (WoS) or Scopus. Citation networks can then be explored interactively, for instance by drilling down into a network and by identifying clusters of closely related publications[5].

The VOS-viewer is a software instrument for constructing and visualizing (mapping) a broad range of bibliometric networks. These networks may for instance include journals, researchers, or individual publications, and they can be constructed with co-citation, bibliographic coupling, or co-authorship relations. In particular, the VOSviewer also offers a text mining functionality that can be used to construct and visualize conceptual (co-word based) networks of terms extracted from a body of scientific literature, particularly titles and abstracts of publications. The VOS viewer can be uploaded with any type of relational information and particularly with publications records of the WoS as well as of Scopus[6].

### *Citation links of SBs*

We analyze the citation network for each set of SBs of the three main fields, i.e., the 389 physics, 265 chemistry, and 367 engineering & computer science SBs as defined earlier in this paper ($\boldsymbol{s}$=10, $\boldsymbol{cs}_{max}$=1.0, $\boldsymbol{a}_{min}$= $\boldsymbol{a}_{max}$=10, $\boldsymbol{ca}_{min}$=5, short notation [10, 1.0, 10, 5]). This was done by creating a full record (title, abstract, authors, institutions, references) set of these SBs[7] and uploading this set into the *CitNetExplorer*. Thus, the SBs are the source publications and the references of the SBs define the citation links. This procedure renders a citation network based on the references of the SBs if sufficient citations links are available. In the visualization of the citation network each circle represents a publication. Publications are labeled by the last name of the first author. To avoid overlapping labels, some labels may not be displayed. The horizontal location of a publication is determined by its citation relations with other publications. The vertical location of a publication is determined by its publication year. The lines represent citation relations, citations point in upward direction: the cited publication is always located above the citing publication. Publications are clustered based on their citation relations. The identified clusters have different colors.

[5] More about CitNetExplorer see http://www.citnetexplorer.nl/Home

[6] More about VOSviewer see http://www.vosviewer.com/Home

We determined the citation network of the 389 physics, 265 chemistry and 367 engineering & computer science SBs. The CitNetExplorer algorithm applies threshold values of important parameters for the construction of the citation network, particularly for the minimum number of citation links and also for the minimum cluster size. In this sense, the CitNetExplorer operates as a community detection tool. We refer to the methodology section in the CitNetExplorer website for details (see footnote 5). A high value for the minimum number of citation links (e.g., 10) results in a sparse network and a low value (e.g., 2) gives an overcrowded picture.

The interactive facilities of the CitNetExplorer allow to find an optimal network configuration. By trying out several parameter values, we find a sensible representation of the overall citation network analysis with 5 for the minimum number of citation links and 2 for the minimum cluster size. A minimum number of 5 citation links means that in the set of 389 physics SBs only references (publications cited by the SBs) that occur at least in 5 different SBs are included in the construction of the network. This provides us with a general overview. The next step is a 'drill down' to specific SBs which reveals more details of the citation network.

First the general overview. The result for the 389 physics SBs is shown in Fig.6a. Several clusters are detected, indicated by colors. Most of these clusters are small, mainly because of the relatively high threshold for the minimum number of citations links. Major clusters are the blue one (left side of the figure) about supergravity and related theoretical work, the gray one about photometric work in biophysics, and the green cluster about string theory and dark matter. In this network we chose a number of SBs as examples for further analysis. These SBs, marked in Fig.6a with a square, are the papers of Ward *et al* (1982), van Neerven and Vermaseren (1984), Romans (1986a), Kostelecky and Samuel (1990), and Kosowsky *et al* (1992). In Table 7 we see that Kostelecky with co-author Samuels have 6 SBs, Romans has 5 SBs, Ward has 3 SBs, and Kosowosky with co-author Turner have 3 SBs. Not all of these SBs are visible in Fig.6a because of the threshold for the number of citation links as discussed above.

[7] Our search algorithm identifies SBs with the requested variables and creates an Excel-database of these SBs including their WoS UT (unique tag) codes. These UT codes can then be uploaded into the WoS website menu in order to produce the full records of the SBs.

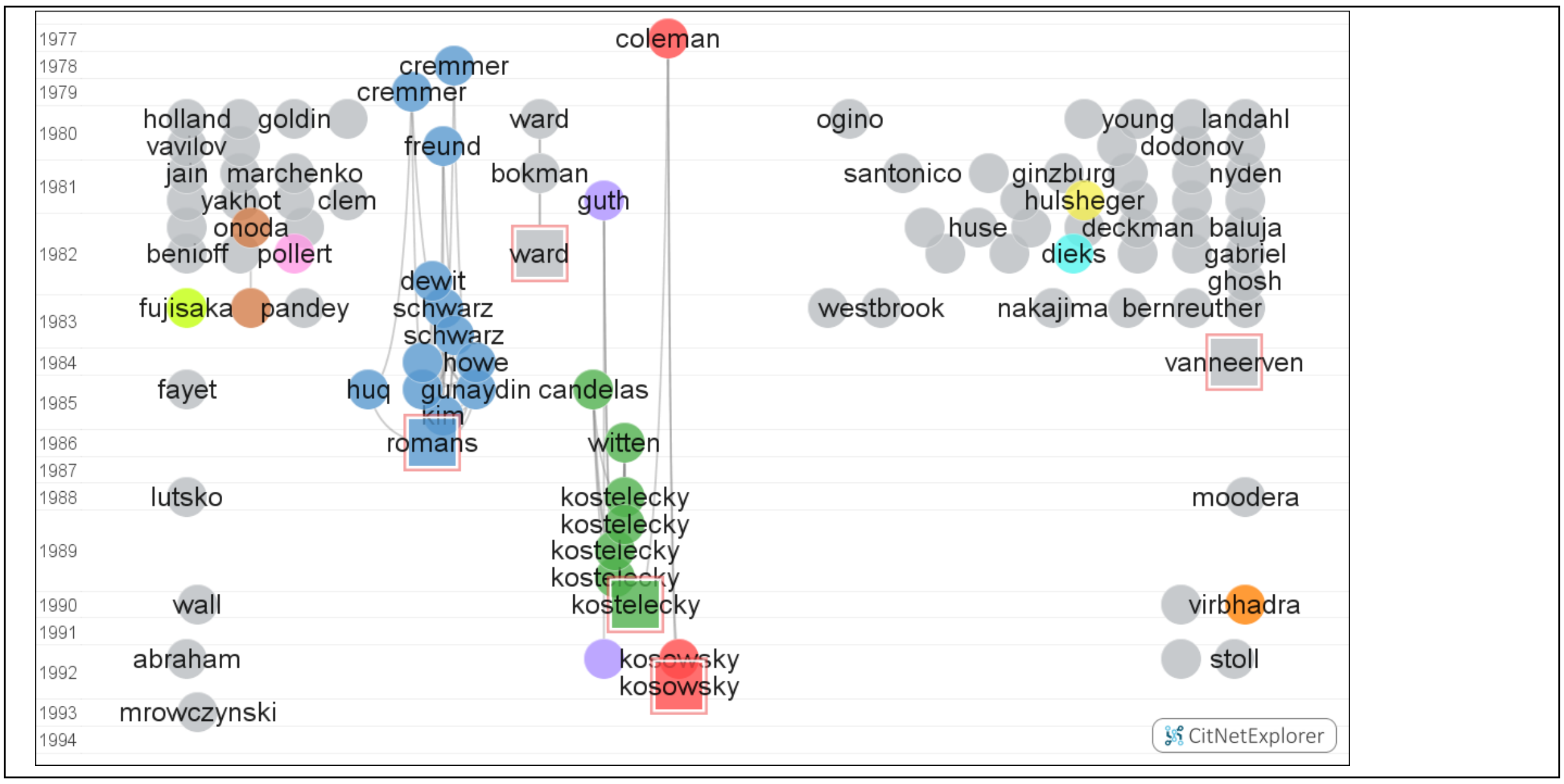

*Figure 6a: Physics Sleeping Beauties and their citation links, overall picture (minimum number of citation links = 5).*

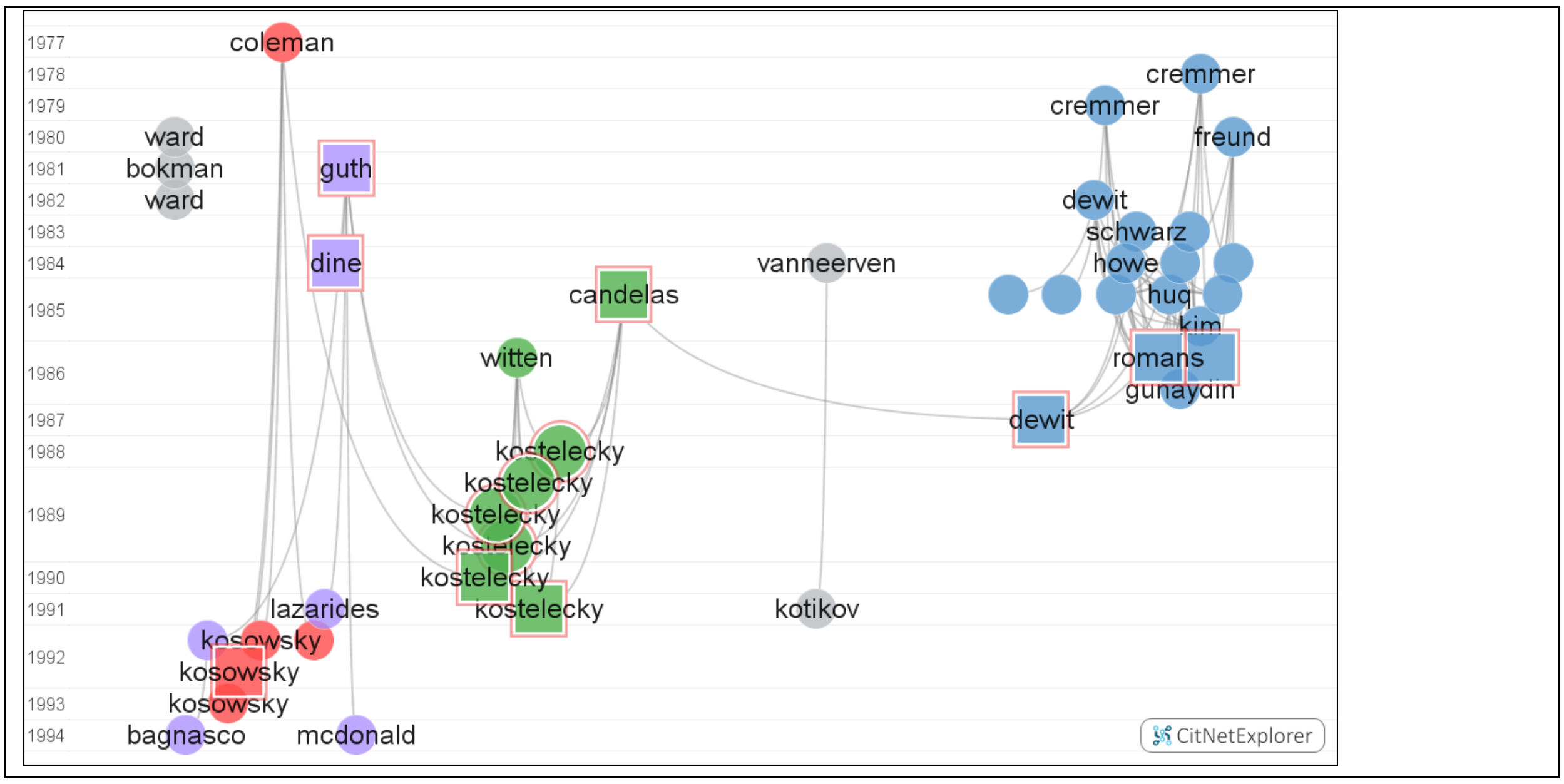

*Figure 6b: Five selected physics Sleeping Beauties clusters and their citation links (minimum number of citation links = 5).*

With the 'drill down' option of the CitNetExplorer the citation network of specifically these selected SBs is extended. We show the results in Fig. 6b. Remarkably, three clusters are clearly connected. From left to right we see the Kosowsky, Kostelecky and Romans cluster. The Kosowsky cluster consists of SBs (of which three by Kosowski and colleagues) on astrophysical, particularly cosmological topics such as dark matter, gravitational radiation and the origin of matter in the inflationary universe. In the Kostelecky cluster we find SBs (of which six by Kostelecky and colleagues) as well as the highly cited paper of Witten (1986) (not a SB, appears in the cluster as a cited publication with a number of citation links above the threshold of 5). This cluster deals with theoretical physics topics on the nature of matter and gravitation, in particular string theory. The Kosowsky and Kostelecky clusters are connected by the SB of Dine *et al* (1984) on supersymmetry (a model on the classification of elementary particles) and the very highly cited paper (not a SB) of Guth *et al* (1981) on the inflationary universe. Via the also very highly cited paper (not a SB) on vacuum configurations for superstrings by Candelas *et al* (1985), co-authored by the string theory pioneer Witten, and via the SB of De Wit *et al* (1987) on supergravity we observe a further connection with the Romans cluster, particularly the most famous SB of Romans (1986b) on supergravitation in a ten-dimensional space-time (van Raan 2004). The SB of McDonald (1994) on the extension of the standard model to include dark matter particles is somewhat isolated between the Kosowsky and the Kostelecky cluster. This a most interesting paper as we shall see further on because this paper becomes highly cited and turns out to have a remarkable form of 'self-awakening'.

In between the above clusters we find the small cluster of Van Neerven and Vermaseren (1984) and Kotikov (1991), both SBs which deal with mathematical methods for describing the interaction between elementary particles. Although these SBs are not connected with the above discussed clusters (at least not above the threshold value for citation links), the position of these SBs is certainly sensible given the importance of mathematical methods in theoretical physics. In the upper left corner of the figure we find the Ward cluster (Ward *et al* 1980, Bokman and Ward 1981, Ward *et al* 1982), all are SBs dealing with biophysical and particularly photochemical topics.

By lowering the threshold for the minimum number of citation links from 5 to 3, more references of the SBs are covered and thus the citation network shows further details. We select the McDonald (1994) SB and drill down again for this SB. The result is given in Fig.6c. Here we see a detailed citation network with links to the seminal books of Weinberg (1972) on gravitation and cosmology and of Hawking and Ellis (1973) on the large scale structure of space-time. The Kosowsky (left hand side of the figure), Kostelecky (middle) and Romans (right) clusters remain connected. Interesting is the separate cluster at the right hand side of the two Death SBs. They are part of three successive papers (in a row in the same journal volume) on gravitation radiation of colliding black hole of which the first one is not an SB (Death and Payne 1992a,b,c). Most probably, researchers working on this topic decide to cite only the first paper and leaving the two next papers uncited, thus creating 'artificially' SBs by a kind of citation-superfluity.

From the above discussed observations we conclude that the physics SBs with theoretical topics are relatively strongly connected along different paths. What about the connectivity of the application-oriented physics SBs? We find that only the Ward cluster has a some connectivity as shown in Figs.6a and 6b. The lowest positioned publication is the SB of Ward et al (1982). The network shows the reference links to earlier publications of Ward and colleagues, and all are SBs. All other application-oriented physics SBs appear to have a very limited amount of citation links. Next to the Ward cluster we can only identify (see Fig.6a) the link between the SBs of Onoda (1982) and Nomura (1983)[8] as a small cluster. This cluster concerns research on the application of ceramics in microwave devices.

[8] The Nomura (1983) SB is positioned in Fig.6a between the SBs marked as Fujisaka and Pandey (Fujisaka and Yamada 1983; Pandey and Mehta 1983).

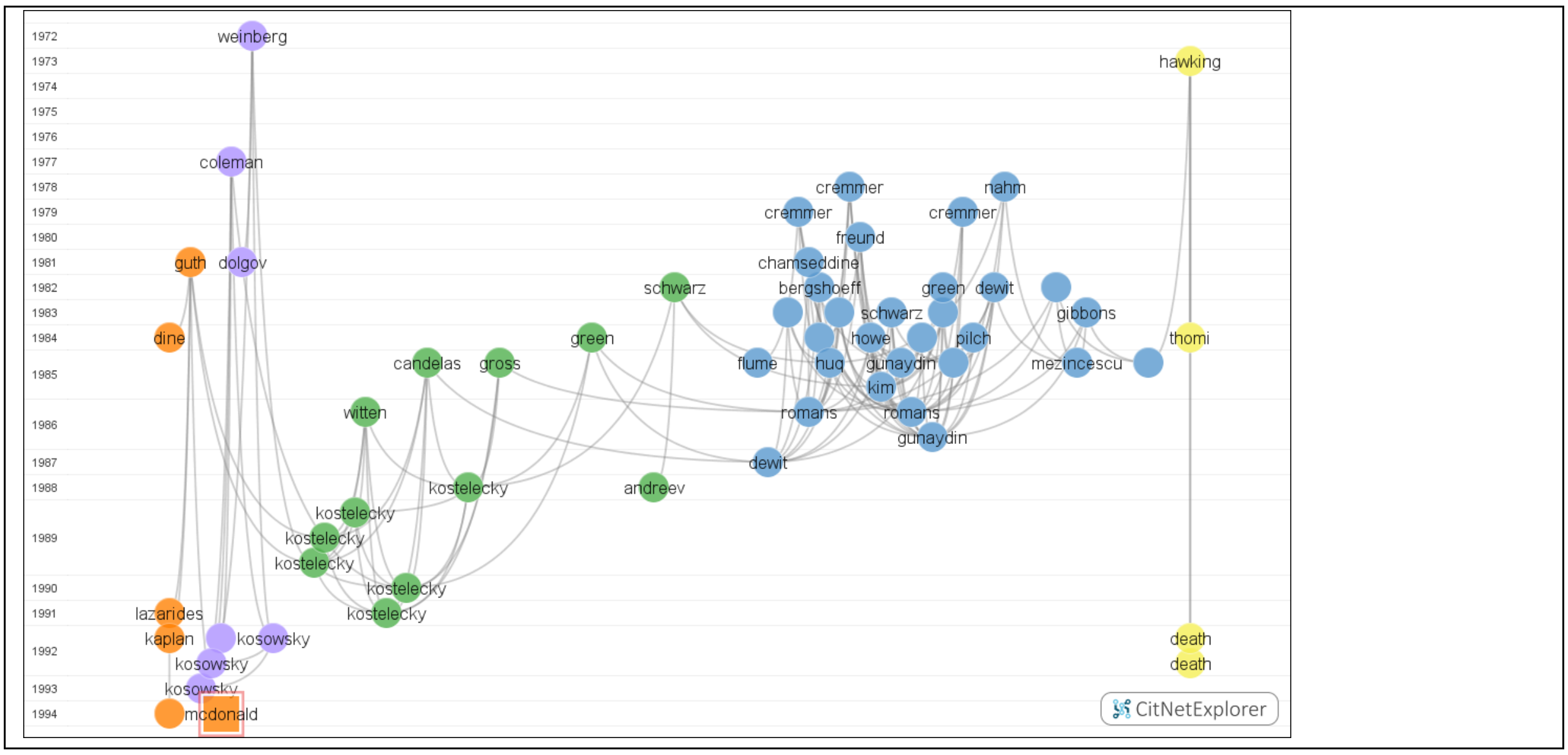

*Figure 6c: Further selection of physics Sleeping Beauties clusters and their citation links, now with lower threshold (minimum number of citation links = 3) and drill down to the McDonald SB.*

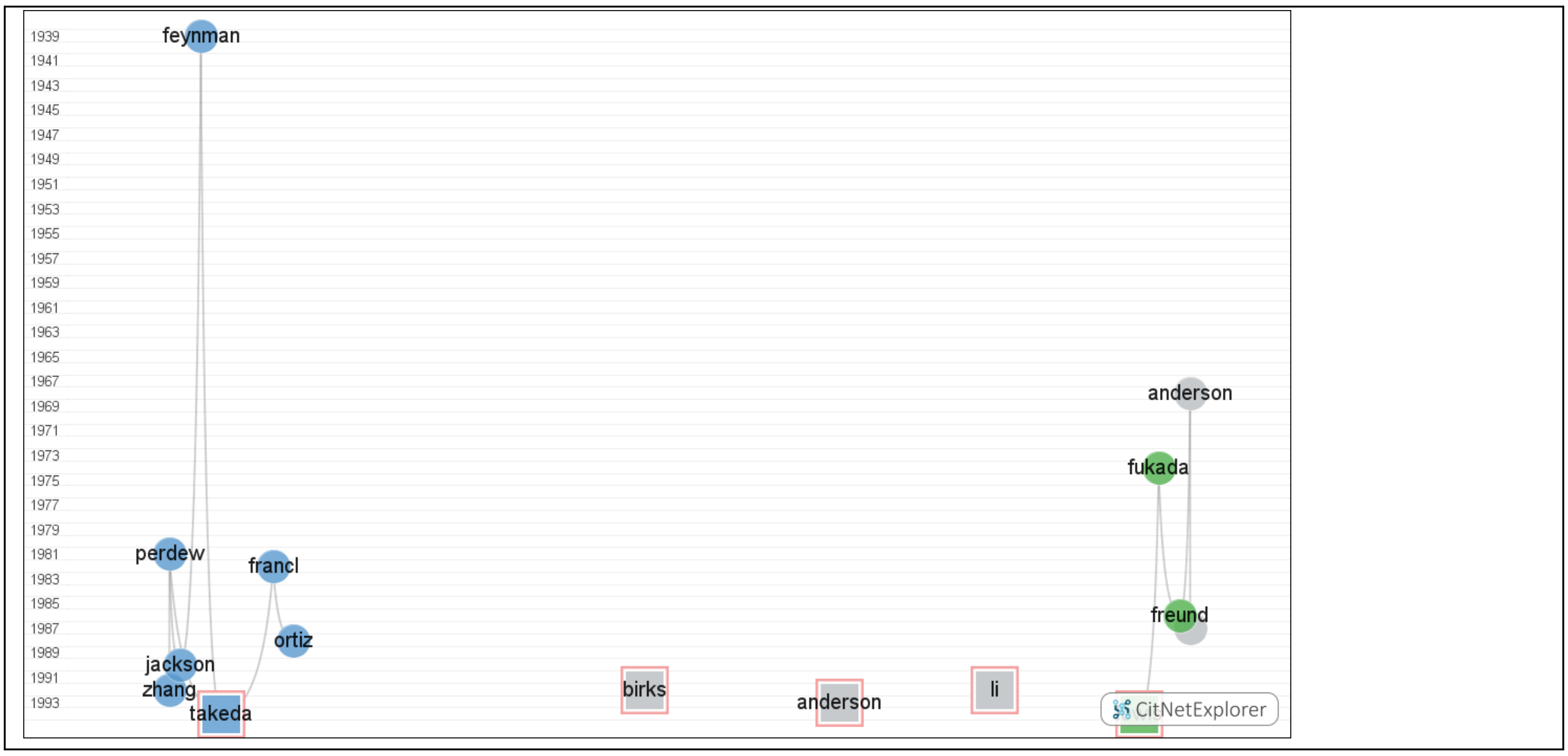

*Figure 6d: Application-oriented physics Sleeping Beauties and their citation links. The Lewis SB is just behind the CitNetExplorer logo (minimum number of citation links = 2).*

To illustrate the rather isolated positions of application-oriented SBs further, we selected from the set of 389 physics SBs those published in the most recent years of this set, 1992, 1993, 1994, and form this subset the 10 most cited (after awakening), of which five are application-oriented. This selection is further discussed in the next section on characteristics of the papers citing the awakened SBs. We present in Fig.6d the citation links map of these five highly cited application-oriented SBs: Birks (1992) on optical fibers; Li and Ahmadi (1992) on particles in turbulent flows; Anderson et al (1993) on propagation of light pulses in nonlinear optical fibers; the earlier discussed SB of Takeda and Shiraishi (1994) on flat silicene structures; and Lewis (1994) on electrical insulation at nanoscale. We see that only the Takeda and Lewis SBs have some citation links above the threshold but none of the application-oriented SBs has direct or indirect a connection with another application-oriented SB.

First results for the citation networks for the chemistry and the engineering & computer science SBs are presented in Fig.A6. We observe less citation links between the SBs than in physics, in agreement with the finding that compared to physics in both main fields a larger part of the SBs is application-oriented.

### *Concept Maps of Sleeping Beauties*

The same WoS-based full record sets of the 389 physics, 265 chemistry and 367 engineering & computer science SBs that are used as source publications for uploading in the CitNetExplorer can also be used for uploading in the VOSviewer. Several choices can be made with the VOSviewer: Co-citation, bibliographic coupling, co-authorship, and term co-occurrence (co-word) networks. We first apply the term co-occurrence facility to create concept maps. After uploading the set of full records, the VOSviewer applies a natural language text processing technique to collects terms (mainly noun phrases) from the titles and the abstracts of the SB publication records. In a next step the VOSviewer calculates all term co-occurrences (i.e., in how many publications of the set any possible pair of terms co-occurs, thus the construction of the co-occurrence matrix) after a specific occurrence threshold has been chosen (interactively with the VOSviewer menu, for instance, a term must occur at least 3 times in the whole set of SBs).

A major challenge in the construction of concept maps is the selection of terms. Although many irrelevant (the, and, of, between, etc.) terms are automatically removed by the VOSviewer natural language text processing algorithm, still the algorithm selects terms such as 'theory', 'approximation', 'dependence', 'correlation', 'possibility', 'calculation', 'comparison', 'increment', 'assumption', 'principle', 'water', 'atom', 'molecule', 'ion', etc. The problem is that these terms may be relevant in some sets of publications, whereas in other sets they are not. Therefore it is sometimes unavoidable to remove specific terms manually. The VOSviewer provides this facility of manual term selection. This, however, is a tricky question. If the set of publications is not large, like in our case of 386 SBs, the removal of just one term, for instance 'principle' may quite drastically change the structure of the map. On the other hand, if rather general terms such as 'principle' are not removed, it is possible that two clusters representing quite distant parts of physics, such as supergravity and crystal structure are 'linked together' because in both subfields the term 'principle' may have a high occurrence (and thus co-occurrence with other terms). Often one has to find the best solution by trial and error, and particularly in the case of smaller sets of publications the choice for full counting instead of binary counting[9] of terms might be a decisive factor.

[9] The VOSviewer offers the possibility to choose for full counting, i.e., all occurrences of a term in a publication are counted; or binary counting, i.e., only presence or absence of a term in a publication is counted, thus the

We here discuss the results for the 389 physics SBs. The results for chemistry and engineering & computer science are presented in the separate file 'Additional data en results for Chemistry and Engineering'. The text processing of the 389 SBs rendered 3,862 terms, of which 380 (10%) meet the occurrence threshold value 2. The resulting map is shown in Fig.7, upper part. We see a landscape with a variety of topics of the 389 SBs. The terms form clusters which are indicated by colors.

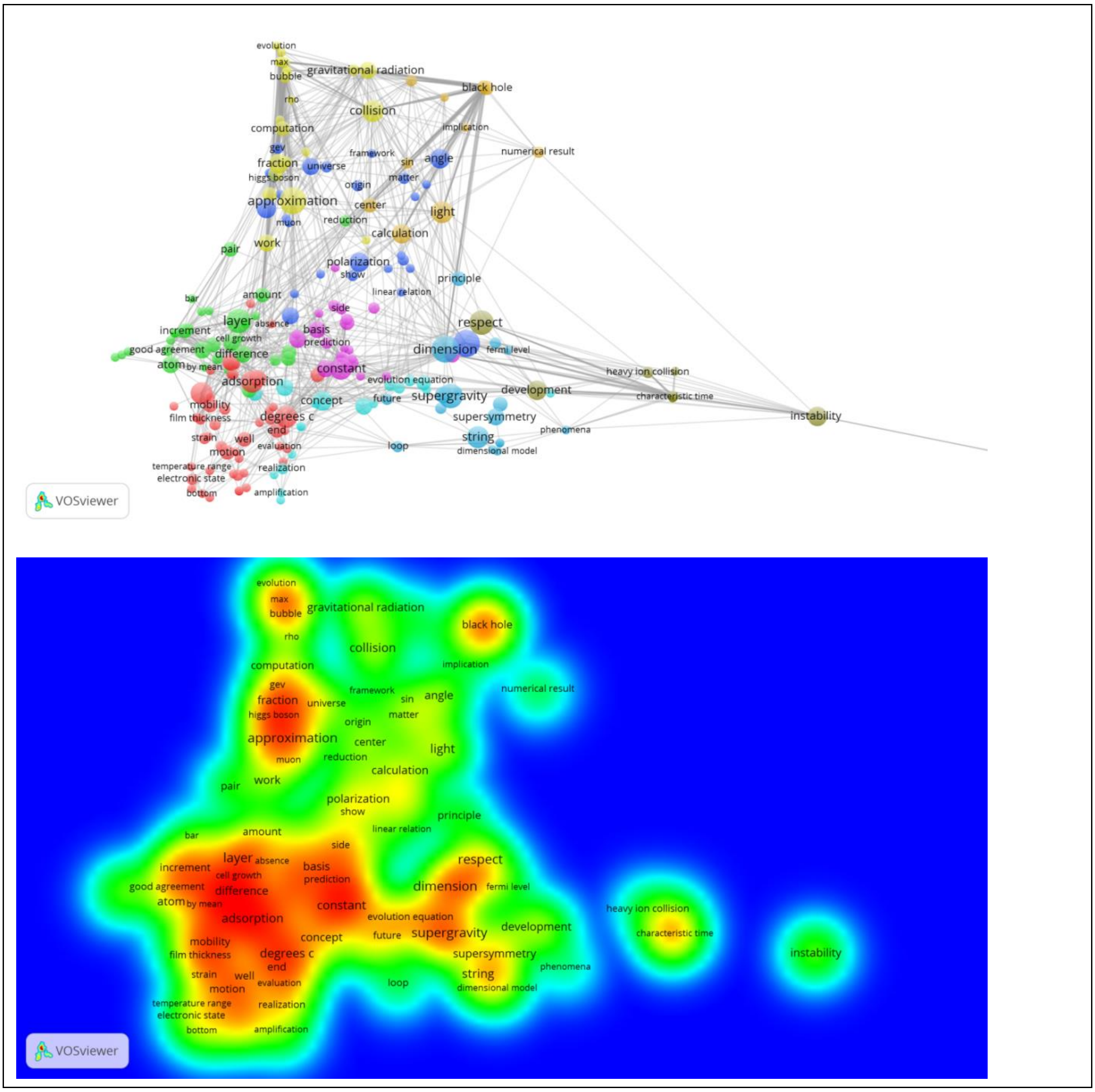

*Figure 7: Concept map of the 389 Physics SBs. Upper part: map with co-occurrence links; lower part: same map, representation of publication densities.*

actual number of occurrences of a term in a publication does not matter. Clearly, in full counting those terms that are mentioned more than once in the abstract of a publication, for instance because these terms are central to the research discussed in the publication, get a heavier weight in the co-occurrence matrix calculations (we refer to http://www.vosviewer.com/Home).

The upper side and the lower right hand side shows the theoretical physics research on gravitational radiation, black holes, supergravity and string theory. Most of the application-oriented work is found on lower left hand side. The same map is also displayed as a density map, see lower part of Fig.7. The more the color shifts from green to red, the higher the number of publications (in this case SBs). In Fig.8, upper part, we zoom into the gravitational radiation cluster. We see important research topics such as black hole, Higgs boson, vacuum bubble, universe. The lower part of Fig.8 is a zoom into the application-oriented part with topics such as luminescence, adsorption, crystal structure, elastic property, refractive index, thin film.

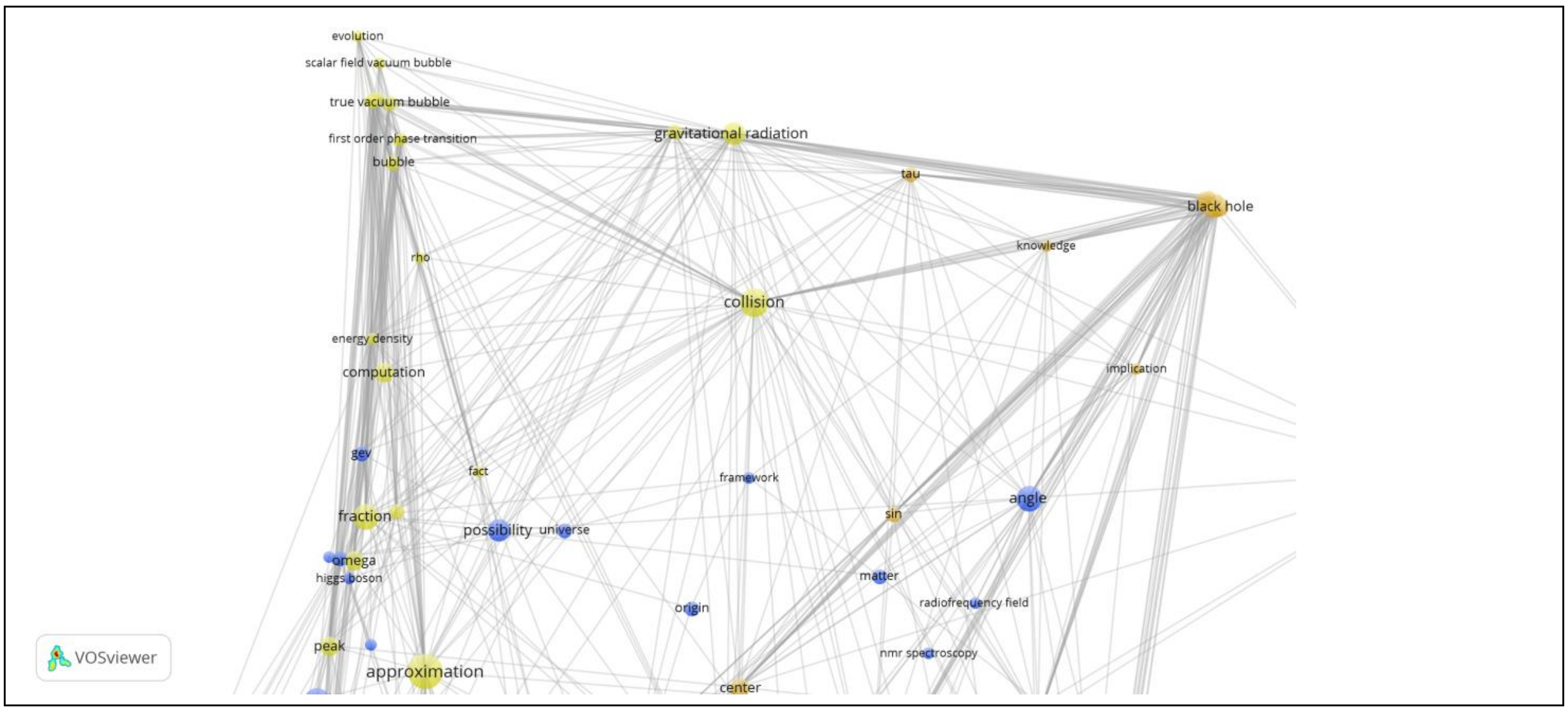

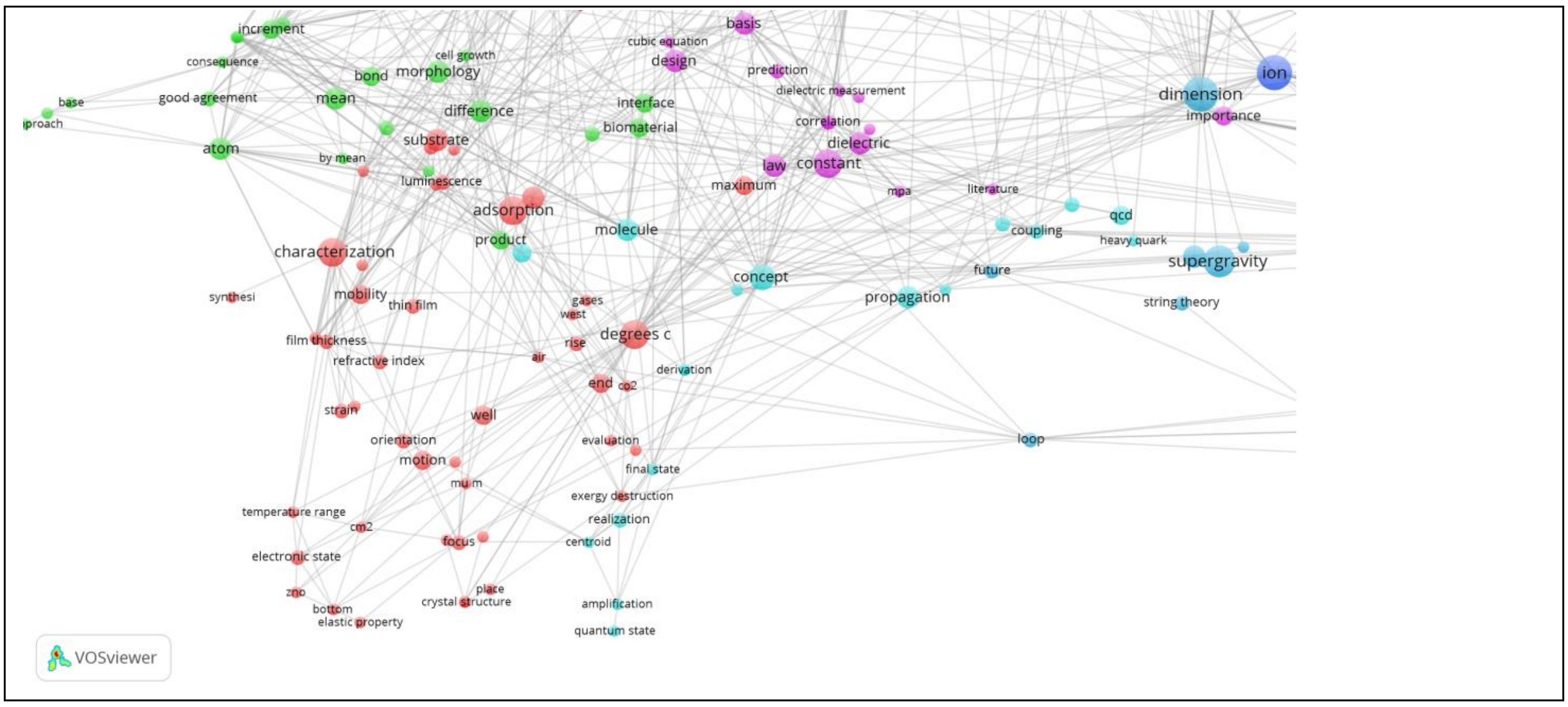

*Figure 8: Zoom into the map shown in Fig.7. Upper part: gravitational radiation cluster; lower part: crystal and thin-layer structure research. Also co-occurrence links are indicated.*

The concept maps for the chemistry and the engineering & computer science SBs are presented in Figs.A7-A8. In chemistry we find for instance clusters related to HIV-polymerase research and to bio-oil research, in engineering & computer science for instance clusters related to algorithms and biomaterials.

### ***Bibliographic Coupling and Co-Citation Maps***

The VOSviewer also offers the possibility to perform a bibliographic coupling (BC) analysis. This means that the SBs are analyzed in their role as *citing* papers and thus a BC-map shows how the SBs are related to each other on the basis of common *cited* papers (the references of the SBs). Fig.9, upper part, shows the results. We observe a cluster of SBs. These are all papers on the theoretical physics of string theory, supersymmetry, supergravity, dark matter and cosmological implications. The SB of Curci (1987) is also on supersymmetry, but it is clearly an outlier because it has references different from the other SBs. It is linked via the SB by De Wit *et al* (1987) with the main cluster.

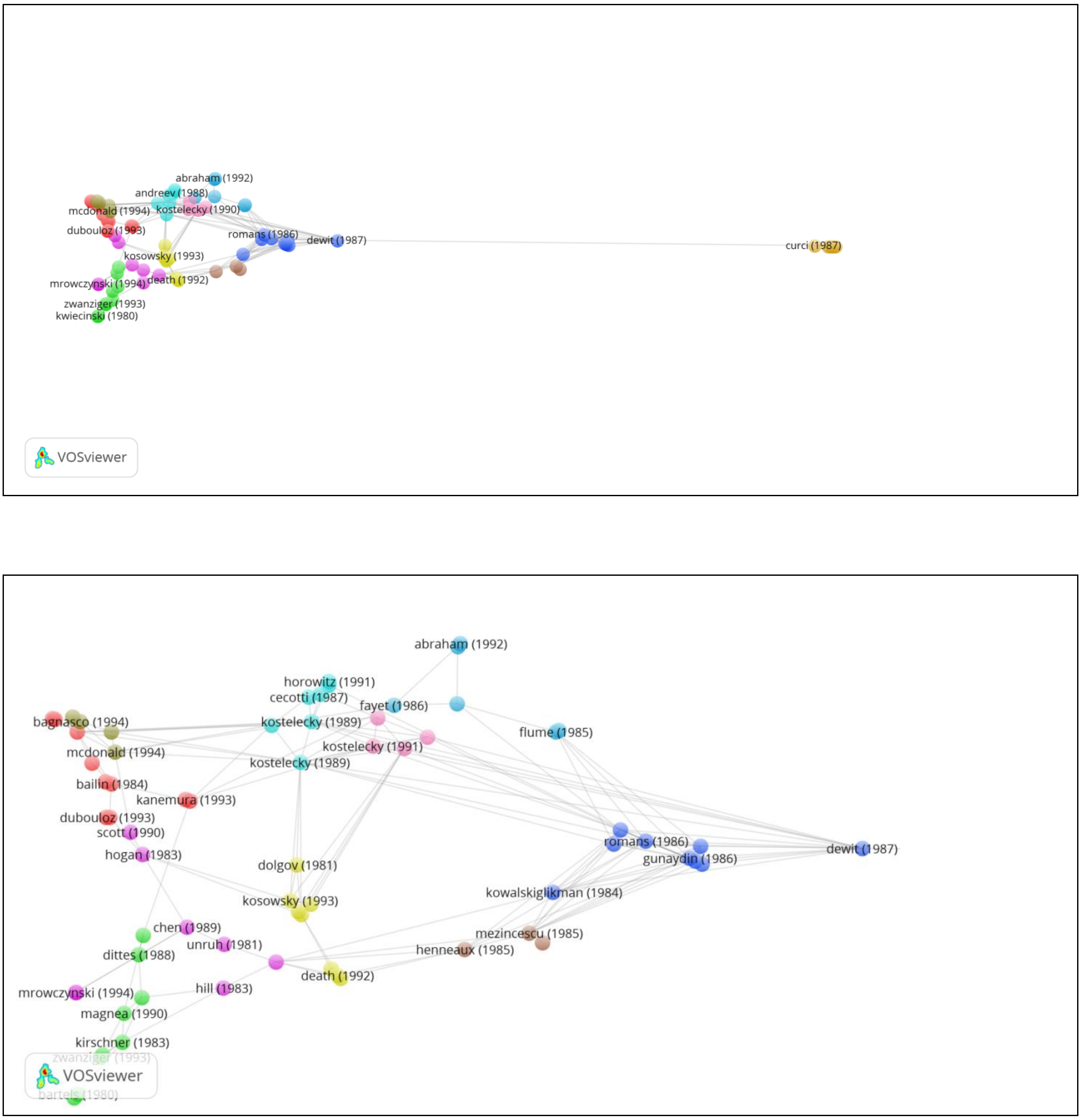

*Figure 9: BC-map of the 389 Physics SBs with links indicating the similarity of cited references. Upper part: overview map; lower part zoom into the main cluster.*

The lower part of Fig.9 shows a zoom into the main cluster, colors indicate different sub-clusters. The SBs of for instance McDonald, Kostelecky, Kosowsky, Romans, Death and De Wit are clearly visible. The BC analysis confirms our findings with the CitNetExplorer: only the theoretical physics SBs are connected with each other, no single application-oriented SB shows up in the BC-map.

In Fig.10 we present the results of the co-citation analysis of the physics SBs. Now the relation of references (*cited* papers) of the SBs (as *citing* papers) are mapped. References can be cited together (co-cited) in a reference list of a paper, and the more this occurs, the stronger their relation. In order to avoid overloading the map, the threshold for the minimum number of cited references is 2. The map of the co-cited references of the physics SBs shows more clusters than in the above discussed case of the bibliographically coupled SBs. Clearly, the references used by the SBs cover a wide range of relevant theoretical physics topics on particles, energy fields, quantum theory, mathematical methods. We see for instance the seminal work of Hawking and Ellis (1973) on the large scale structure of space and time, also identified with the CitNetExplorer. Also in this CC-map only theoretical elementary particles physics papers are visible and no application-oriented papers.

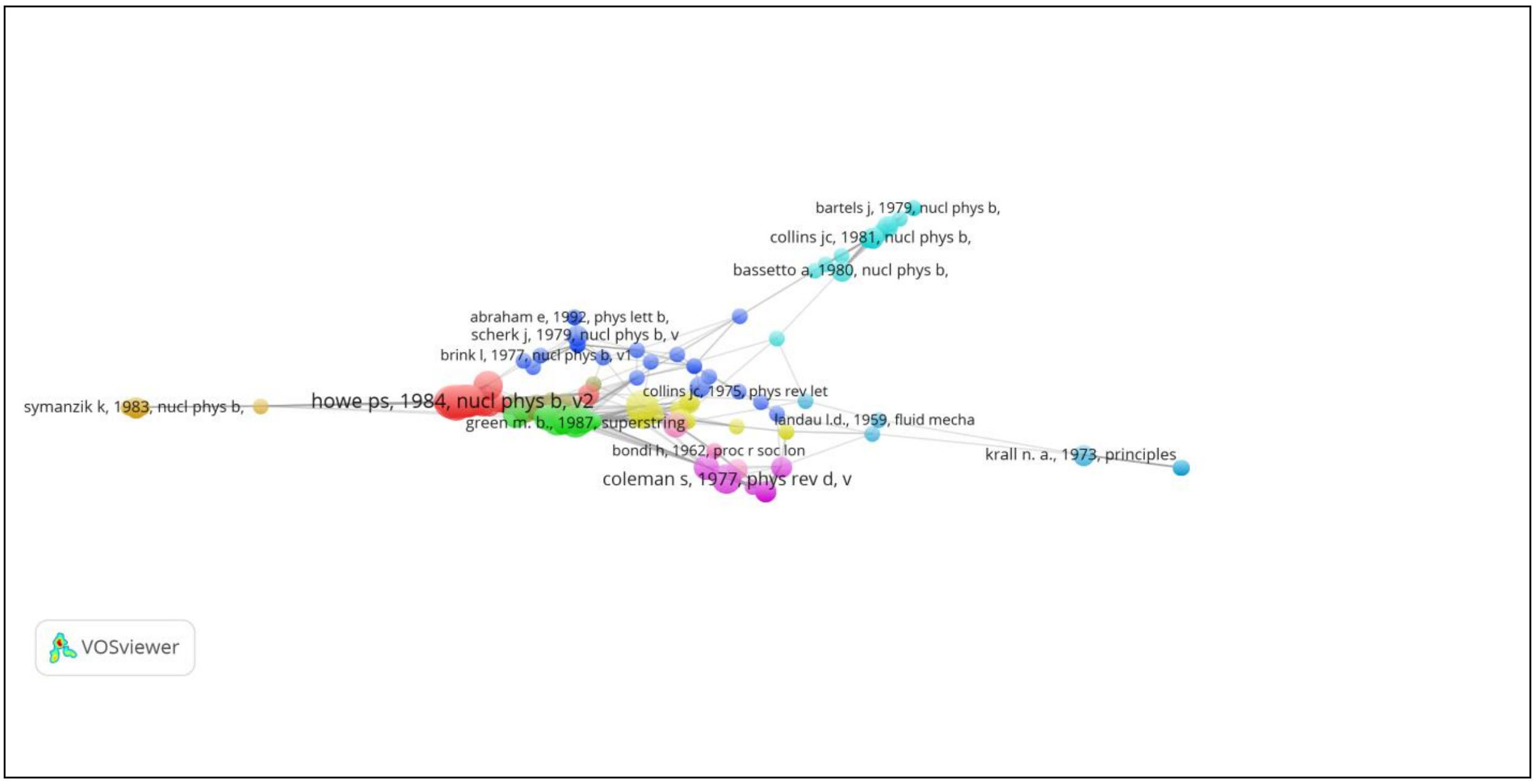

*Figure 10: CC-map of the 389 Physics SBs with links indicating co-citation relations.*

The BC- and CC-maps for chemistry and for engineering & computer science SBs are presented in Figs.A9 and A10, respectively. The BC-map for chemistry shows little structure. We find a cluster of SBs with biological, medical and food research topics such as enzymes. Also a cluster of photochemical research on proteins around the SBs of Ward et al (1980, 1982) is present. The Ward SBs are also physics SBs because of their classification in both biophysics as well as in biochemistry. This photochemical cluster is also present in the CC-map of the chemistry SBs. The BC-map for the engineering & computer science SBs shows little structure too. Small clusters of operational research and communication technology are present. In the CC-map we find small clusters related to environmental research and food technology.

## Citing Sleeping Beauties: Characteristics of the Princes

### *Main lines of the approach*

The most important life event of a Sleeping Beauty is her awakening, i.e., the first citation after the sleeping period. But in most cases this is not a well-determined event. For instance, if we allow a certain threshold for the maximum citation rate $\boldsymbol{cs}_{\max}$ during the sleeping period -as in the case of the 389 physics SBs discussed in this paper where $\boldsymbol{cs}_{\max}$=1.0- one could say that every now and then a prince comes along without much effects. Only in the very rare cases of 'coma sleep', $\boldsymbol{cs}_{\max}$=0, there is no attention at all – in terms of citations- for the SB.

The 'real' awakening of the SB is mostly not a 'one prince only' action. This nicely illustrated by Fig.5. In these examples, only in the case of the Romans SB there is one specific citing paper after 10 years in 1995. This case is discussed in Van Raan (2004) were it was shown that this first citing paper appeared in fact in 1996 but the volume of the journal was formally registered as the last volume of 1995. Then, in 1996 we see already 11 further citing papers. In the example of Simon (1989) the awakening year is 2001, 12 years after publication, with 4 citing papers. In the case of Dieks (1988), an SB that became, and still is, highly cited, the awakening year is 1998, 10 years after publication, with 6 citing papers. Thus, the awakening of an SB often reflects a development which is, be it quite suddenly, 'in the air' resulting in several citing papers followed by more and more citing papers.

Why was the SB awakened? A first step to answer this question is to find characteristics of the citing publications. A next and decisive step is interviewing one or several 'princes' (authors of the first-year-after-sleep citing papers) of each SB separately. In this paper we will focus on the first step and as a follow-up we are preparing the interviews.

One possibility to find out whether an SB is indeed work ahead of time, would be a comparison of the concept-maps of the 389 physics SBs with concept-maps of the entire physics literature in the time period that the SBs were published (1980-1994). But this approach proved to be very cumbersome because the specific topics of SBs simply disappear in the enormous amount of other publications in the same years. In a follow-up study we will confine the entire physics literature to publications closely related to the topics of the SBs. This may enhance the probability to analyze to what extent the topics of an SB deviate from the main stream in its own research field.

In this paper we take another approach. We narrowed down the scope and selected from the 389 physics SBs the most recent ones, namely those published in 1992, 1993, 1994, in total 122 SBs. The year 1994 is the last year because we need a total time span of 20 years. Next, we selected from these 122 SBs the most cited (after the sleeping period, until May 2015). These top-10 physics SBs have in total 286 references (publications cited by these SBs) and 2,464 citing publications. In Table 8 we present the details of these top-10 physics SBs together with a short description of the importance of the paper which is in fact the reason why it was awaked. We notice that also in this small set 50% is application-oriented, namely the papers of (first author) Birks, Li, Takeda, Anderson and Lewis.

AHARONOV Y, DAVIDOVICH L, ZAGURY N (1993). QUANTUM RANDOM-WALKS. *PHYSICAL REVIEW A* 48( 2) 1687-1690.
*This work coins for the first time the term 'quantum random walk'. Has become important since the development of quantum computers.*

BIRKS TA, LI YW (1992). THE SHAPE OF FIBER TAPERS. *JOURNAL OF LIGHTWAVE TECHNOLOGY* 10(4) 432-438.
*This work presents a model for the transmission of light through optical fibers. Has become important since the introduction of nano-scale fibers.*

MCDONALD J (1994). GAUGE SINGLET SCALARS AS COLD DARK-MATTER. *PHYSICAL REVIEW D* 50(6) 3637-3649.
*This work describes an extension of the Standard Model of elementary particles to include dark matter. Has become important since the increasingly stronger evidence for the existence of large amounts of dark matter in the universe.*

SCHOLTZ FG, GEYER HB, HAHNE FJW (1992). QUASI-HERMITIAN OPERATORS IN QUANTUM-MECHANICS AND THE VARIATIONAL PRINCIPLE. *ANNALS OF PHYSICS* 213(1) 74-101.
*This works describes a new mathematical approach to define physical observable quantities within the quantum mechanics framework. Has become important in various fields of physics such as nuclear physics and superconductivity research.*

GONZALEZ J, GUINEA F, VOZMEDIANO MAH (1994). NON-FERMI LIQUID BEHAVIOR OF ELECTRONS IN THE HALF-FILLED HONEYCOMB LATTICE - A RENORMALIZATION-GROUP APPROACH. *NUCLEAR PHYSICS B* 424(3) 595-618.
*This work describes the dielectric behavior of 2D atomic structures. Has become important since the development of graphene sheets.*

LI A, AHMADI G (1992). DISPERSION AND DEPOSITION OF SPHERICAL-PARTICLES FROM POINT SOURCES IN A TURBULENT CHANNEL FLOW. *AEROSOL SCIENCE AND TECHNOLOGY* 16(4) 209-226.
*This work describes the behavior of small particles in turbulent flows. Has become important for medical applications of colloids and in fluid dynamics of the respiratory system.*

TAKEDA K, SHIRAISHI K (1994). THEORETICAL POSSIBILITY OF STAGE CORRUGATION IN SI AND GE ANALOGS OF GRAPHITE. *PHYSICAL REVIEW B* 50(20) 14916-14922.
*This work presents a theoretical model of a flat hexagonal atomic structure of silicon. Has become important since the development of silicene and graphene sheets.*

ANDERSON D, DESAIX M, KARLSSON M, LISAK M, QUIROGATEIXEIRO ML (1993). WAVE-BREAKING-FREE PULSES IN NONLINEAR-OPTICAL FIBERS. *JOURNAL OF THE OPTICAL SOCIETY OF AMERICA B-OPTICAL PHYSICS* 10(7) 1185-1190.
*This work describes the propagation of light pulses through nonlinear fibers. Has become important because nonlinear optical fibers are essential to drastically improve telecommunication infrastructure.*

LEWIS TJ (1994). NANOMETRIC DIELECTRICS. *IEEE TRANSACTIONS ON DIELECTRICS AND ELECTRICAL INSULATION* 1(5) 812-825.
*This work coins the term nanometric dielectrics for the effect of nano-sized particles on electric insulators. Has become important since the development of, particularly, polymer nanocomposites.*

MROWCZYNSKI S (1993). PLASMA INSTABILITY AT THE INITIAL-STAGE OF ULTRARELATIVISTIC HEAVY-ION COLLISIONS. *PHYSICS LETTERS B* 314(1) 118-121.
*This work focuses on the understanding of one of three basic natural forces, the strong interaction. Has become important, particularly since the CERN announcement in 2000 of evidence for a new state of matter at extremely high densities and temperatures.*

*Table 8: The 10 most cited physics SBs published in 1992-1994.*

### *Concept Map of highly cited papers citing the top-10 Sleeping Beauties*

To create a general overview of the work based, at least partly, on the top-10 SBs, we selected from the whole set of papers citing the top-10 SBs (n= 2,464) the 500 most cited. In other words, top-papers citing top-SBs. The text processing of these 500 citing top-papers rendered 8,810 terms, of which 868 meet the occurrence threshold value 3. We used again the binary counting method (see footnote 9). The resulting map is shown in Fig.11. In this map formed by the 500 citing top-papers we can clearly observe the same research themes as those of the majority of the top-10 SB: quantum walks (Aharonov *et al* 1993), and hamiltonian (Scholtz et al 1992), dark blue cluster; optical fibers and non-linear optical pulse effects, light green cluster (Birks 1992, Anderson 1993); dark matter, red cluster (Mc Donald 1994); particle deposition and turbulence, upper part of purple cluster (Li and Ahmadi 1992); nano-structures, lower part of the purple cluster (Lewis 1994); heavy-ion collisions, light blue cluster (Mrowczynski 1993).

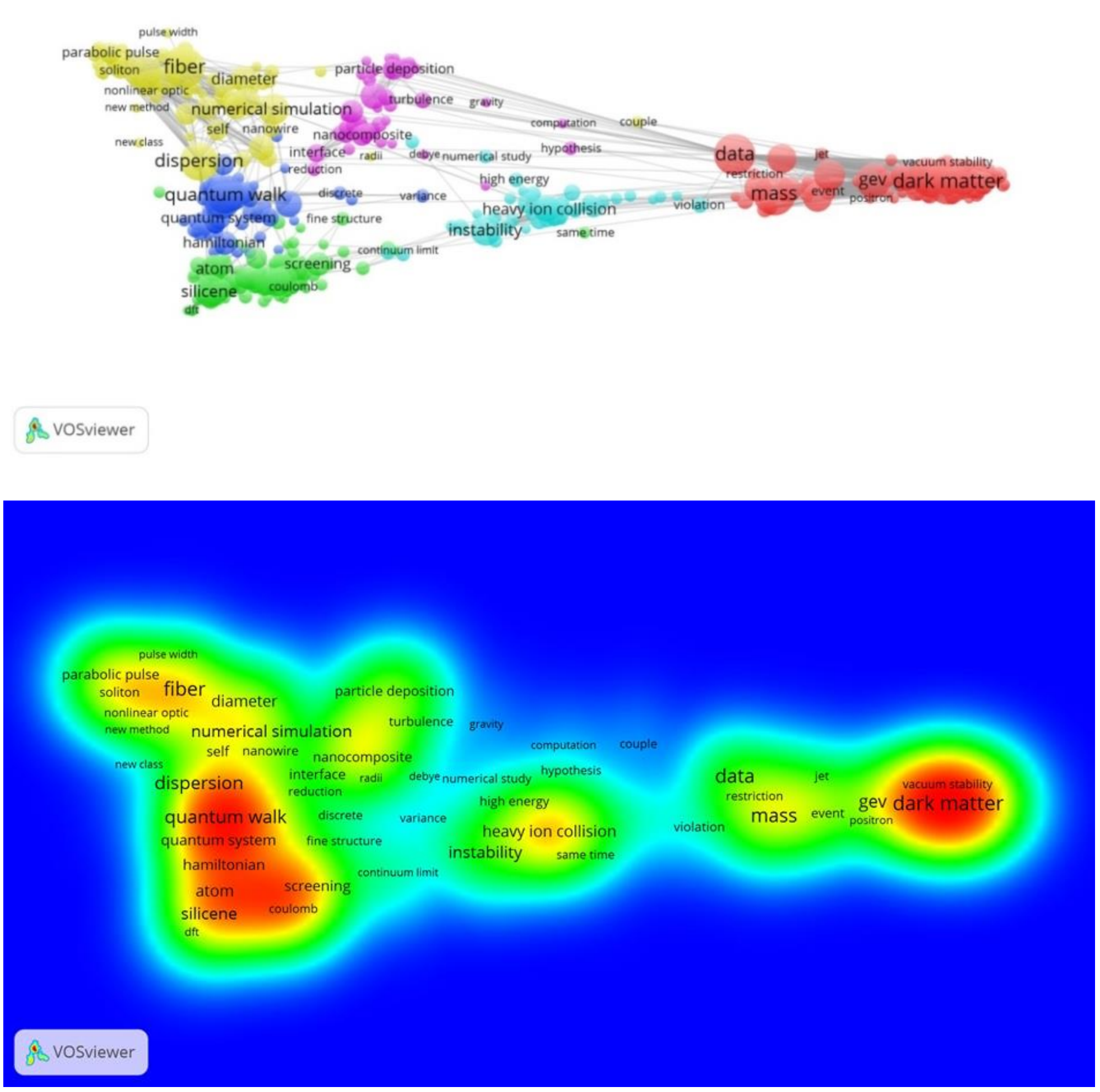

*Figure 11: Concept map of the 500 top-papers that cite the top-10 SBs. Upper part: map with clusters and co-occurrence links; lower part: same map, representation of publication densities.*

The honeycomb lattice of atomic structures (Gonzalez et al 1992) can be found if we zoom into the silicone cluster (green). This silicone cluster is very interesting. In Wikipedia under 'Silicene'[10] we find "Although theorists had speculated about the existence and possible properties of silicene researchers first observed silicon structures that were suggestive of silicene in 2010. …. In 2015, silicene field-effect transistor made its debut that opens up new opportunities for two-dimensional silicon for various fundamental science studies and electronic applications". The first of these 'speculating' theorists is our top-10 Sleeping beauty of Takeda and Shiraishi (1994). This paper is one of the longest sleeping papers (15 years) in recent time, it is a true Sleeping Innovation because it has become very important for the current development of graphene applications.

### *Citation links and concept maps of the SBs: who triggered the awakening?*

First we analyze with the CitNetExplorer the citation relations of the SBs, both the cited papers (the references of SBs) as well as the citing papers (after awakening, a few during the sleeping period, because we allow a maximum citation rate of 1.0 during the sleeping period). As an examples for our analytic procedure, we take 2 Top-SBs, a 'blue skies' SB, McDonald (1994), and an application-oriented SB, Takeda & Shiraishi (1994).

The citation links of the McDonald SB are shown in Fig.12. This SB is cited 330 times until now. In this case the citation analysis clearly reveals the triggering publication (the 'prince'): Burgess et al (2001). This prince was followed a year later by several self-citing papers of McDonald, a kind of self-reinforcing the awakening. However, it took another few years before papers citing the Burgess paper also start to cite the McDonald SB, as clearly visible in the figure. After the SB was awakened and citing papers were already increasing, McDonald further reinforced his own awakening by re-publishing in 2007 his SB of 1994 in arXiv, the repository of electronic preprints in predominantly physics, with an added note "In light of recent interest in minimal extensions of the Standard Model and gauge singlet scalar cold dark matter, we provide an arXiv preprint of the paper, published as Phys.Rev. D 50 (1994) 3637, which presented the first detailed analysis of gauge singlet scalar cold dark matter" (McDonald 2007).

Fig.13 shows the citation links of the Takeda & Shiraishi SB. This SB is cited 238 times until now. Also in this case the citation analysis clearly reveals a prince: Cahangirov *et al* (2009) is the triggering publication. Clearly, the awakening tool a relatively long time. In contrast to the McDonald SB, the number of papers citing the Takeda & Shiraishi SB increases very rapidly immediately after the triggering, as can be seen in the figure.

Certainly not all SBs have such clear triggering publications. In follow-up work we will investigate other, more 'diffuse' types of awakening, in some cases with attempts to self-awakening.

[10] http://en.wikipedia.org/wiki/Silicene, accessed on May 22, 2015.

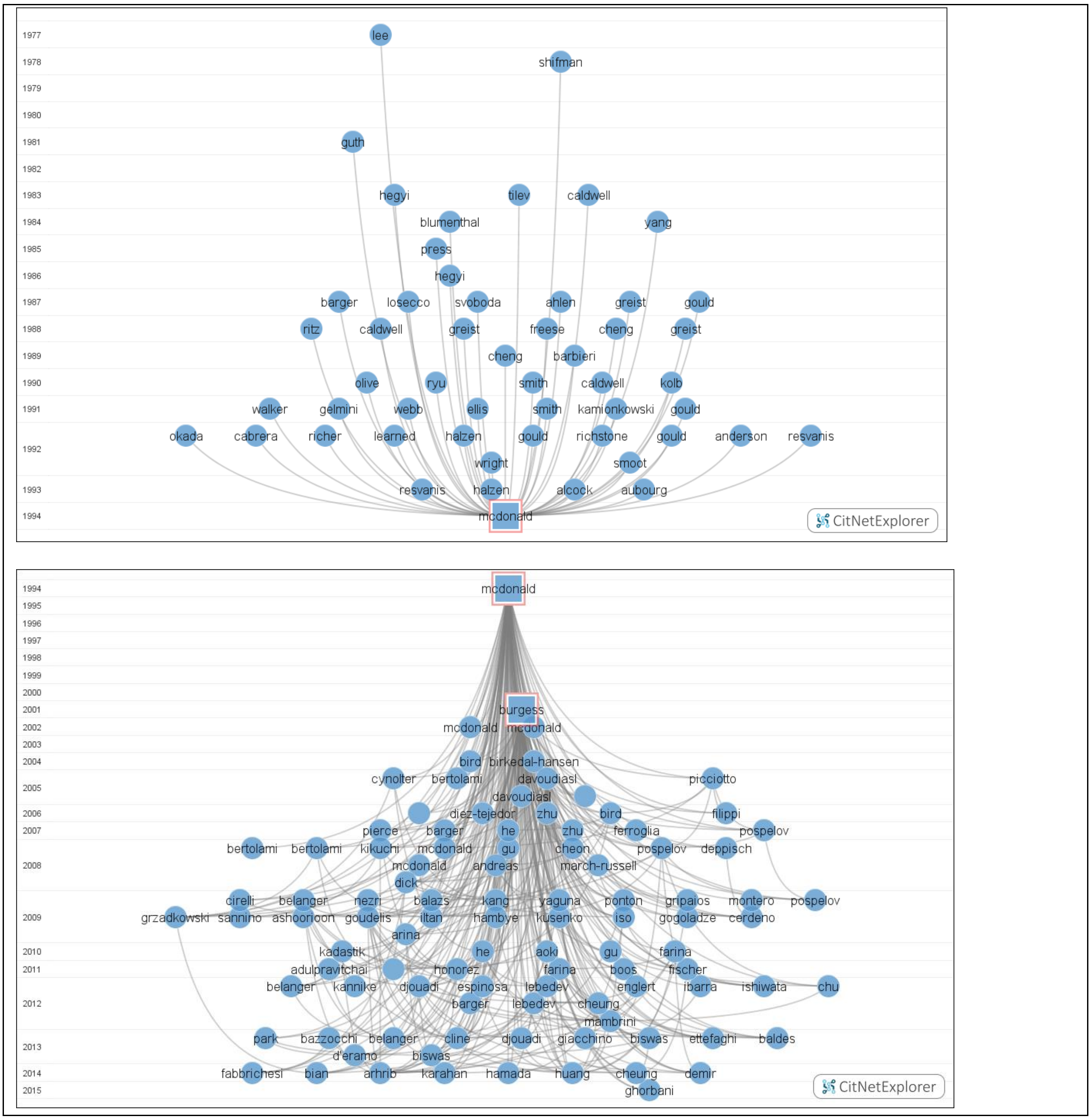

*Figure 12: Cited papers (references) and citing papers of the McDonald Sleeping Beauty. Upper part: cited papers; lower part: citing papers. The time runs top-down.*

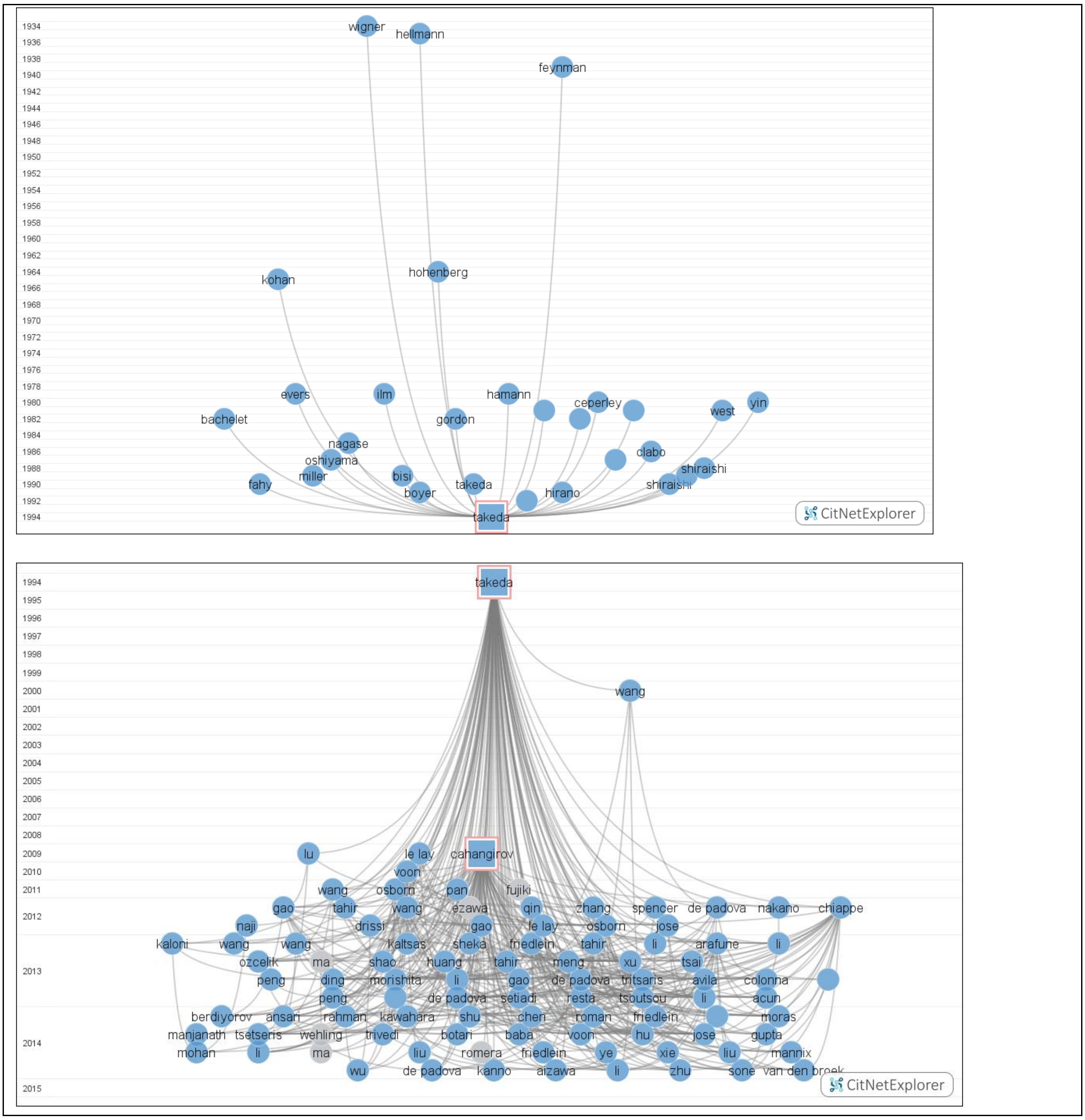

*Figure 13: Cited papers (references) and citing papers of the Takeda & Shiraishi Sleeping Beauty. Upper part: cited papers; lower part: citing papers. The time runs top-down.*

With the VOSviewer we created concept maps of both top-SBs. For the McDonald SB we show (1) the references (n=55) of the McDonald SB in Fig.14, upper part; (2) the McDonald SB itself (n=1) in Fig.14 middle part; and the citing papers of the McDonald SB (n=330) in the lower part of Fig.14. Because of the low number of references, the two first concept maps are inevitably sparse. In particular, the map of the McDonald SB itself does not show any significant structure (all terms are related to the same extent with each other, which causes the ring-shaped structure), but still it shows all relevant terms of this paper.

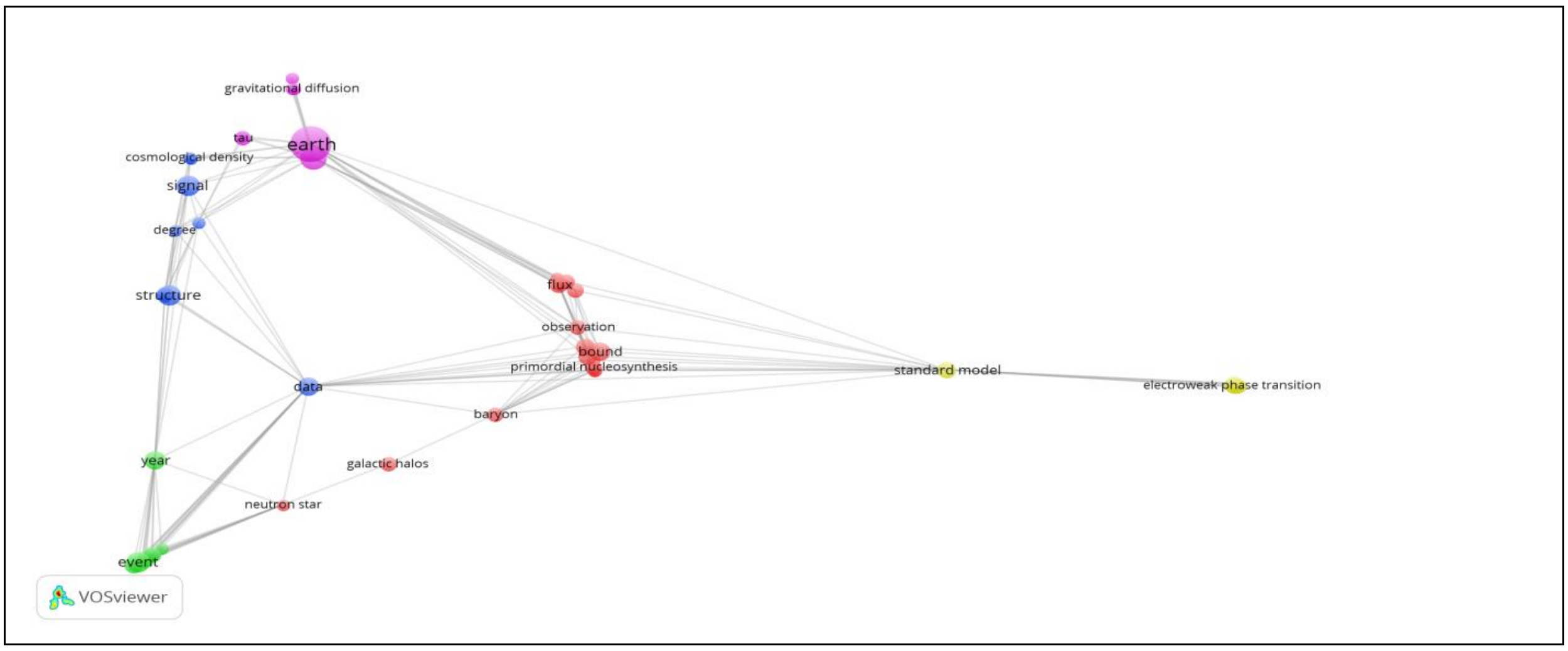

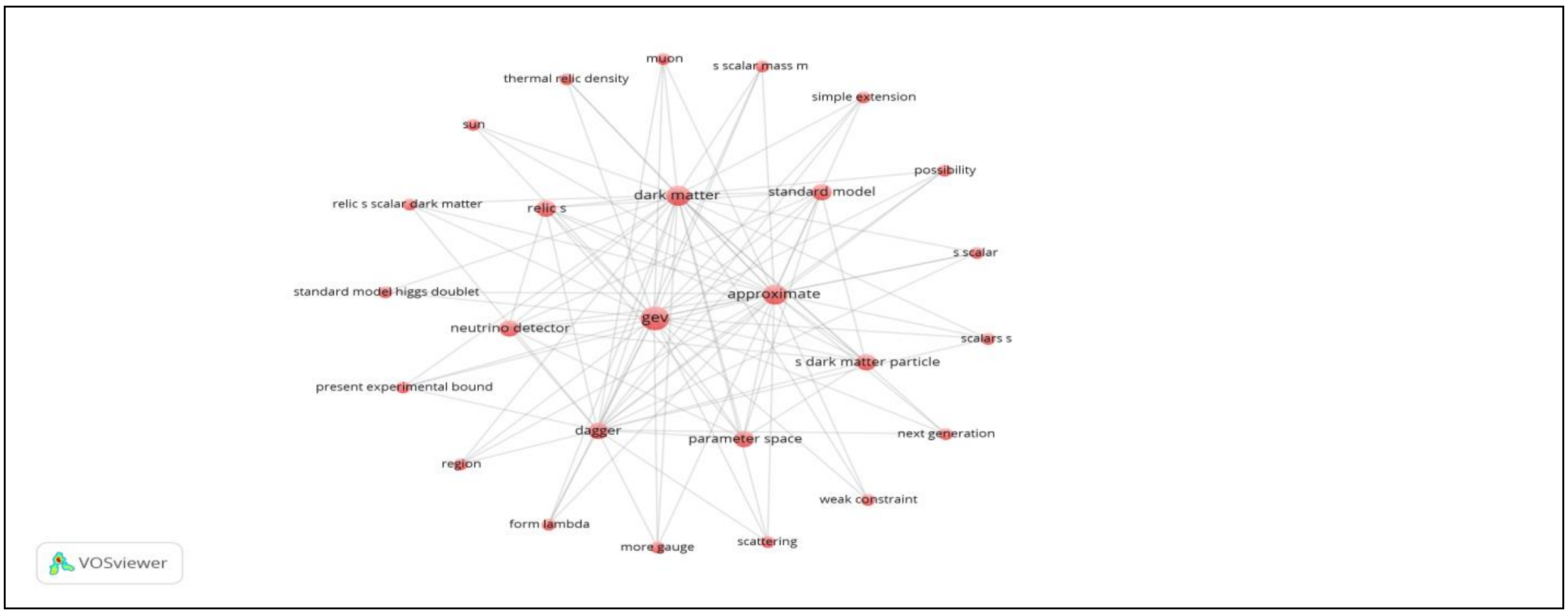

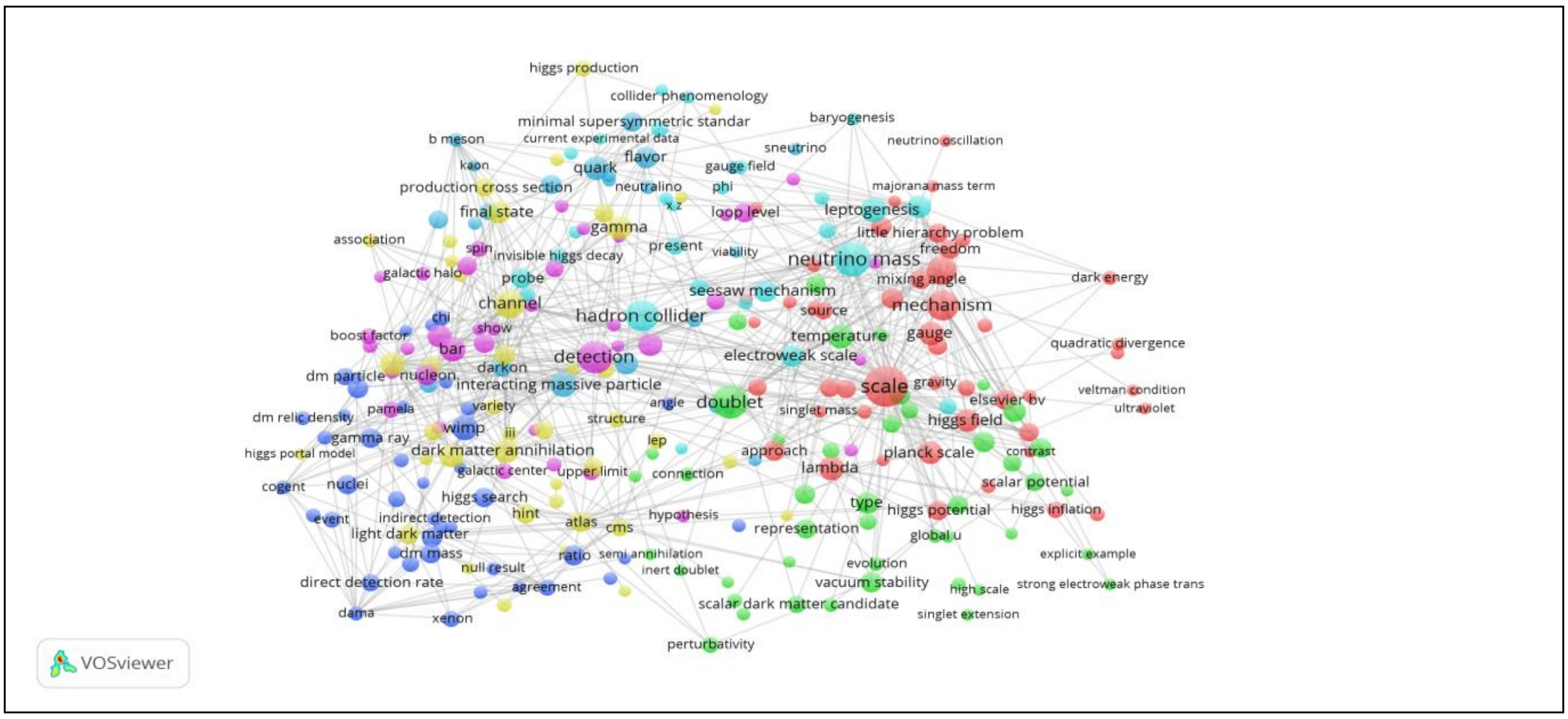

*Figure 14: Concept maps of the references of the McDonald SB, upper part; (2) the McDonald SB itself, middle part; and the citing papers of the McDonald SB, lower part.*

As can be seen, the higher number of citing papers yields a much more extensive concept map. By comparing the three maps we see the evolution of the physics of the universe. As main topics found in the references, most of them from the 1980's and the early 1990's, we see standard model (of the elementary particles), gravitational diffusion, electroweak phase-transition, baryon. Dark matter is clearly not a prominent topic yet. In the map of the McDonald SB dark matter and scalar are the major topics, which is obvious given the title of this paper. The map of the citing papers, almost all are from 2005 until now, shows a landscape of many related topics with recent developments such as the Majorana particle, dark matter candidates, Higgs particle, dark energy. By enhancing the resolution of the concept map also the sterile neutrino (a strong candidate for a dark matter particle) becomes visible, see Fig.15.

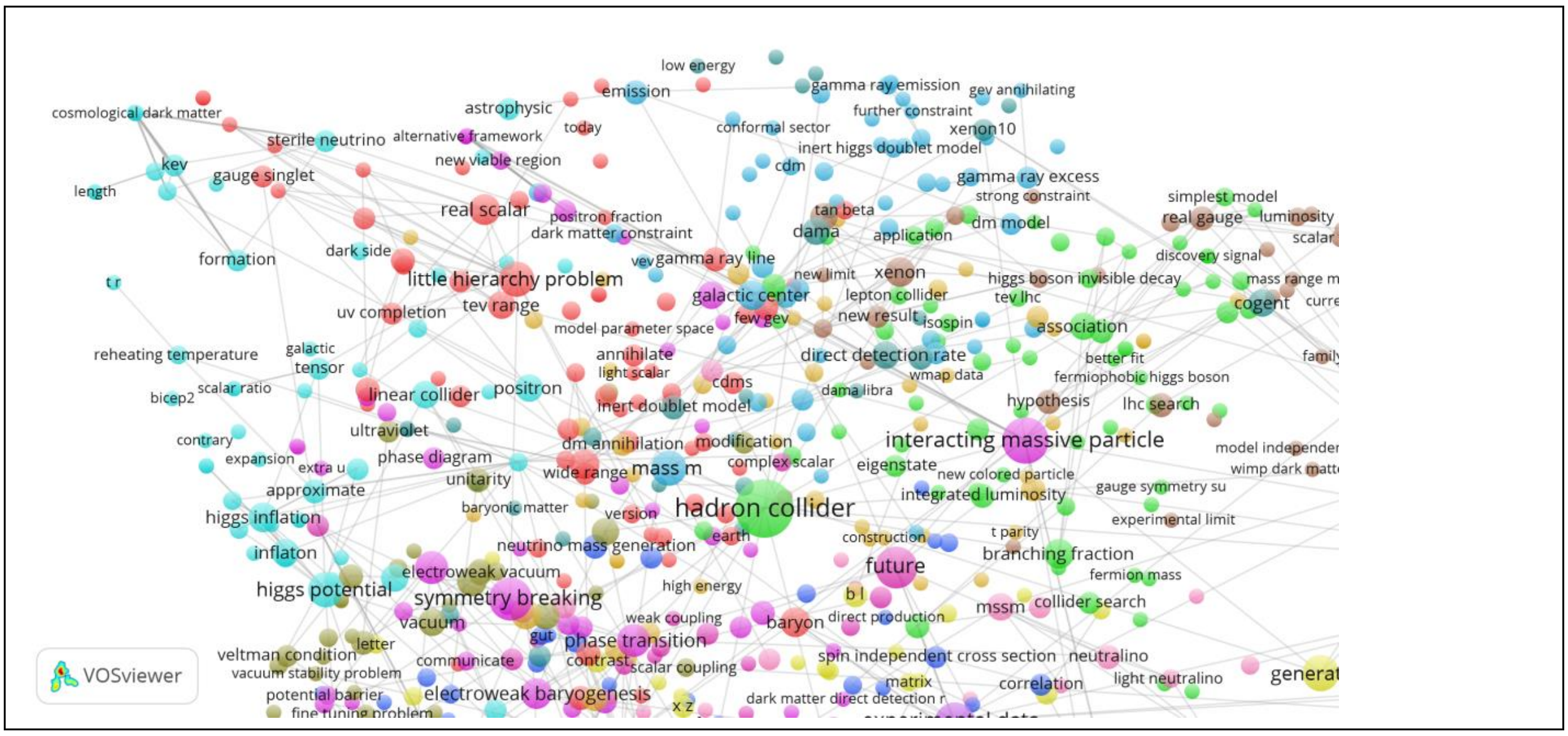

*Figure 15: Concept map of the citing papers of the McDonald SB with higher local resolution, the sterile neutrino is in the upper left corner.*

For the Takeda & Shiraishi SB both the number of items in both the references as well as in the abstract of the paper itself is too small to have meaningful maps. The concept map of the citing papers of Takeda & Shiraishi is shown in Fig.16. The upper part of the figure represent the overview map and the lower part is a zoom into the cluster with silicene nanoribbon and graphene nanoribbon, two central items in the current development of this field of applied physics.

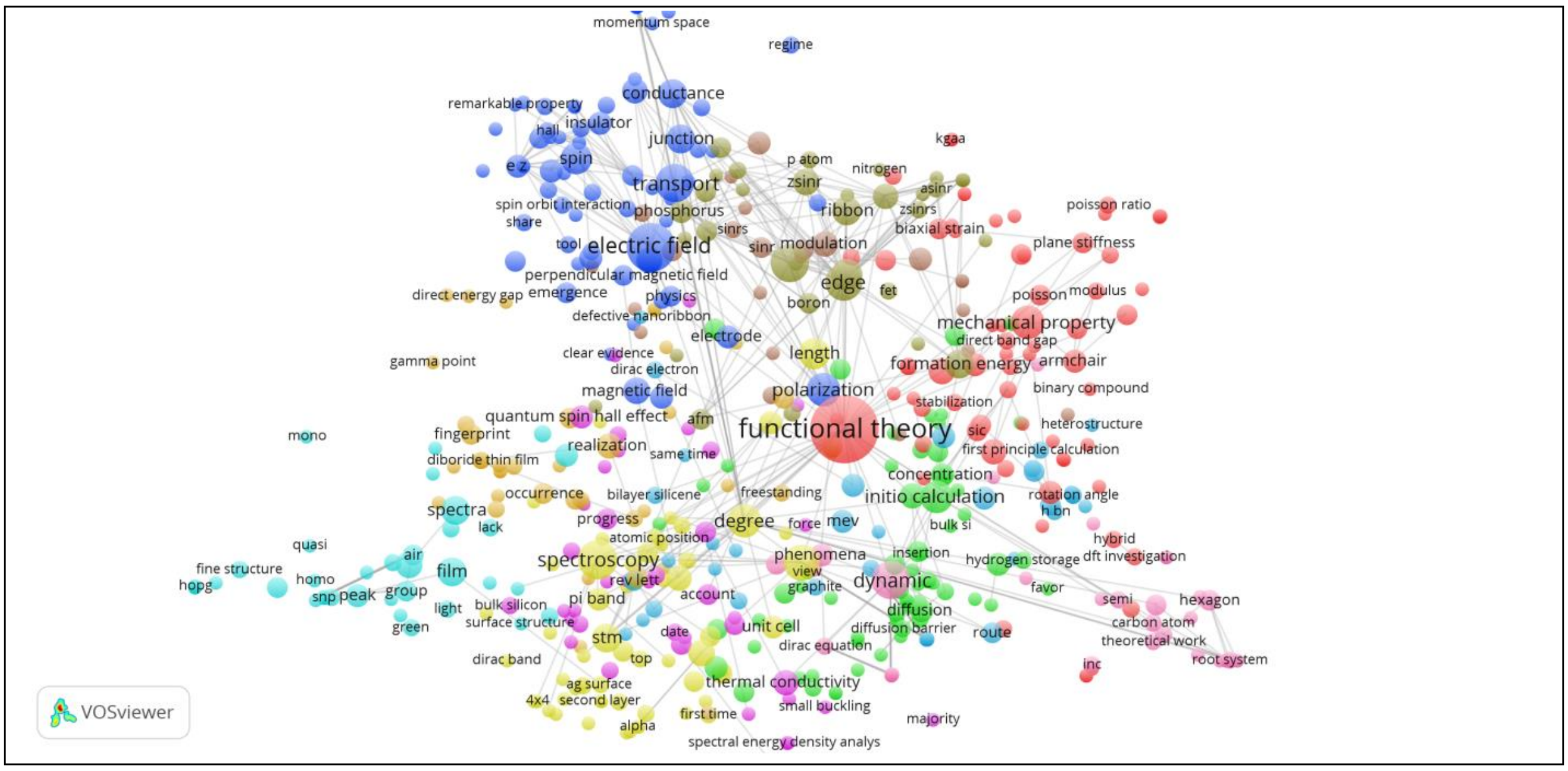

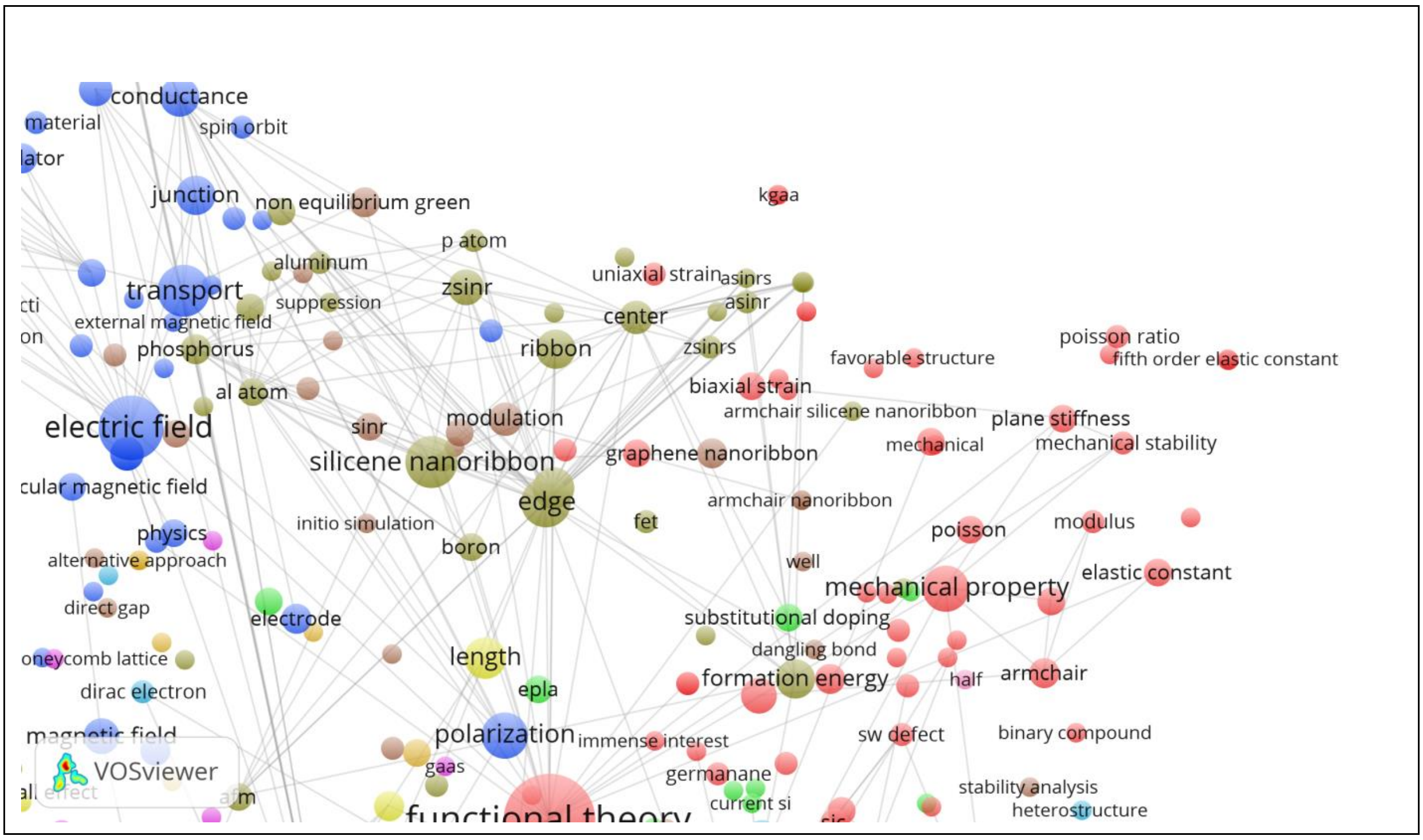

*Figure 16: Concept map of the citing papers of the Takeda & Shiraishi SB. Upper part: overview map; lower part: zoom into silicone and graphene nanoribbon.*

### ***Bibliographic Coupling and Co-Citation Maps***

We also constricted bibliographic coupling and co-citation maps for both Top-SBs. Fig.17 shows the bibliographic coupling (BC) map of the citing papers of the McDonald SB. Thus, these citing papers are clustered and linked on the basis of their common references. In other words, the BC map reveals a picture of the recent literature structured on the basis of their references. We observe different clusters which is to be expected because the citing papers do not only cite the McDonald SB, but many more papers in research themes related to the McDonald work. Nevertheless, because McDonald cites in several of his later papers his own SB, his new work appears in the light blue color in the upper part of the map, together with other authors working on the same or closely related topics.

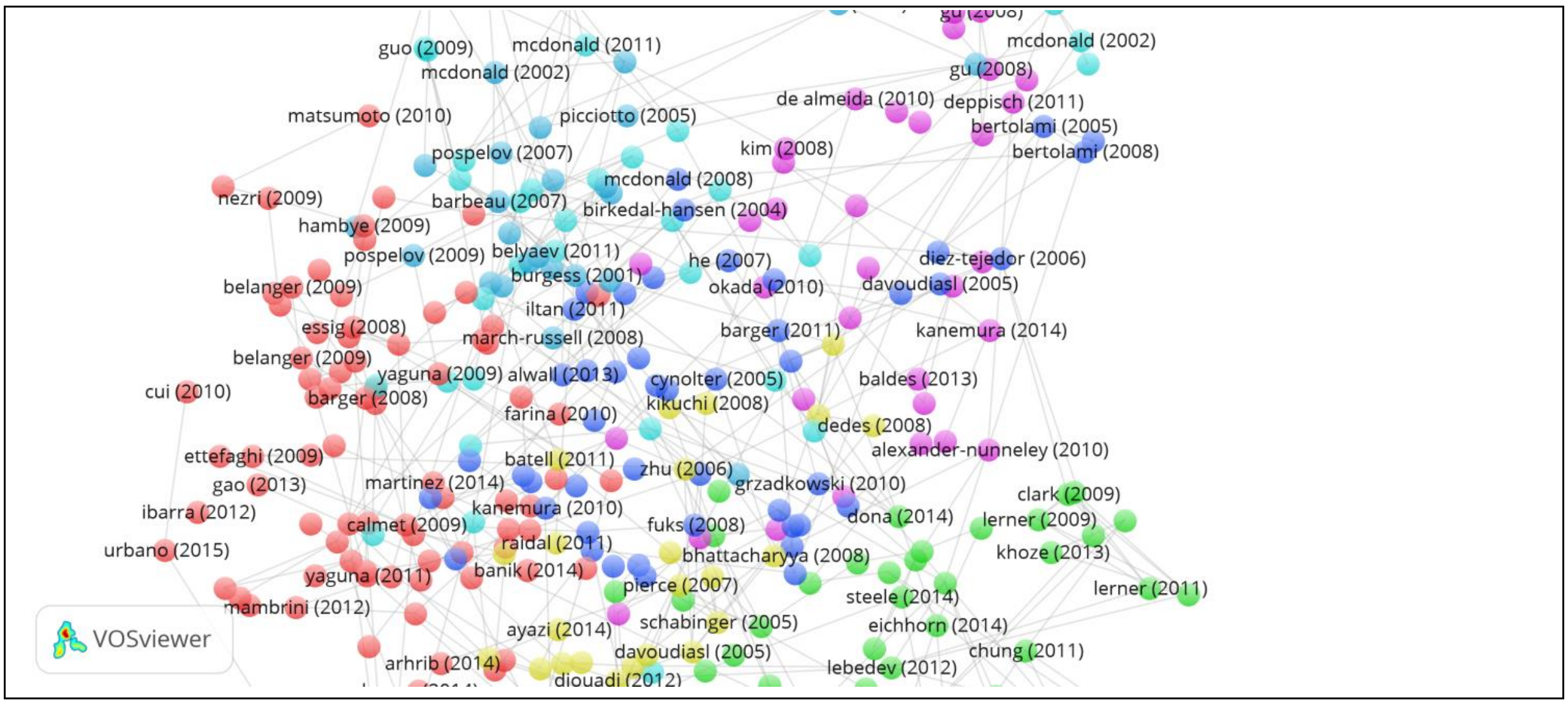

*Figure 17: Bibliographic coupling map of the of the citing papers of the McDonald SB with links indicating the similarity of cited references.*

Fig.18 shows the results of the co-citation analysis of the citing papers of the McDonald SB. Thus, a clustering of the references of the citing papers is constructed, in which the McDonald SB must show up, as all citing papers cite by definition the McDonald SB. This means that the McDonald SB has a prominent position, and this is clearly visible, it is the largest blue circle right in the center of the map. But very close to the McDonald SB we see the paper of Burgess et al (2001) which we discussed in the foregoing section (Fig.12, lower part) as the 'early prince'. As discussed, after the publication of the Burgess paper it took several years before papers citing the Burgess paper also start to cite the McDonald SB. Of the 328 times the McDonald SB is cited, it is 270 times co-cited with the Burgess paper, thus a very strong co-citation relation which is also clearly visible. The co-citation maps also shows that most of the co-cited literature is quite recent: this reflects the habit of researchers (at least in the natural sciences) to cite preferably the recent literature.

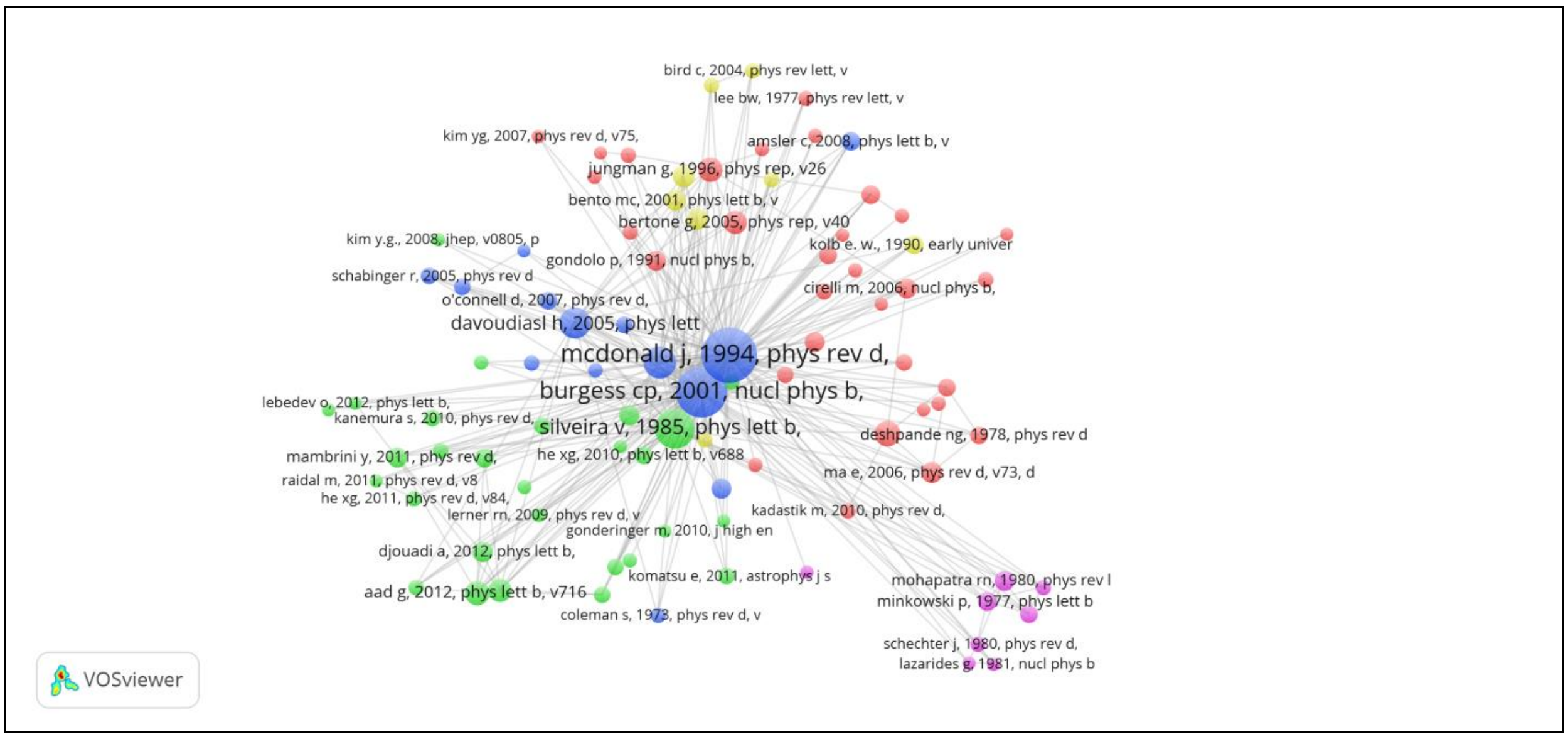

*Figure 18: Co-citation map of the citing papers of the Mc Donald SB with links indicating co-citation relations.*

The bibliographic coupling (BC) map of the citing papers of the Takeda & Shiraishi SB is shown in Fig.19. Thus, these citing papers are clustered and linked on the basis of their common references which provides a landscape of the recent literature in the research themes involved. As in the case of the McDonald SB, we observe different clusters indicating that many papers in different research themes are all related to the Takeda & Shiraishi work.

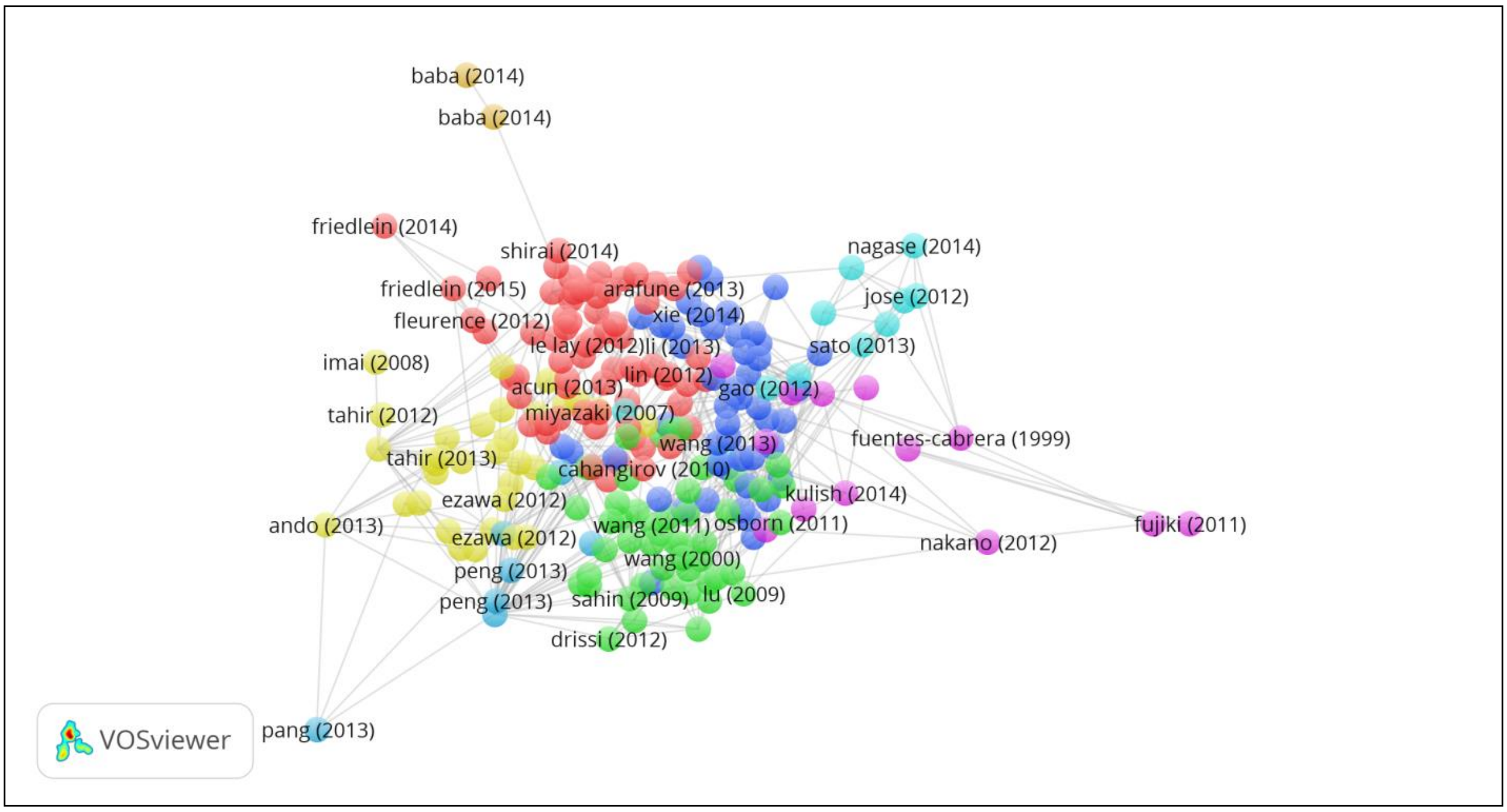

*Figure 19: Bibliographic coupling map of the of the citing papers of the Takeda & Shiraishi SB with links indicating the similarity of cited references.*

Fig.20 shows the results of the co-citation analysis of the citing papers of Takeda & Shiraishi SB. Thus, a clustering of the references of the citing papers is constructed, in which the Takeda & Shiraishi SB must have a central position, as all citing papers cite by definition this SB. As in the case of the McDonald SB, this central position is clearly visible, it is the largest green circle right in the center of the map. Also in this case, the co-citation map clearly shows the publication that triggered the awakening, the prince: very close to the Takeda & Shiraishi paper we see the paper of Cahangirov *et al* (2009). As discussed, in contrast to the triggering of the McDonald SB awakening, the late awakening of the Takeda & Shiraishi SB by the Cahangirov *et al* paper was immediately followed by a strongly increasing number of citing papers up till now. Of the 238 times the Takeda & Shiraishi SB is cited, it is 167 times co-cited with the Cahangirov *et al* paper, so that also here a very strong co-citation relation is clearly visible.

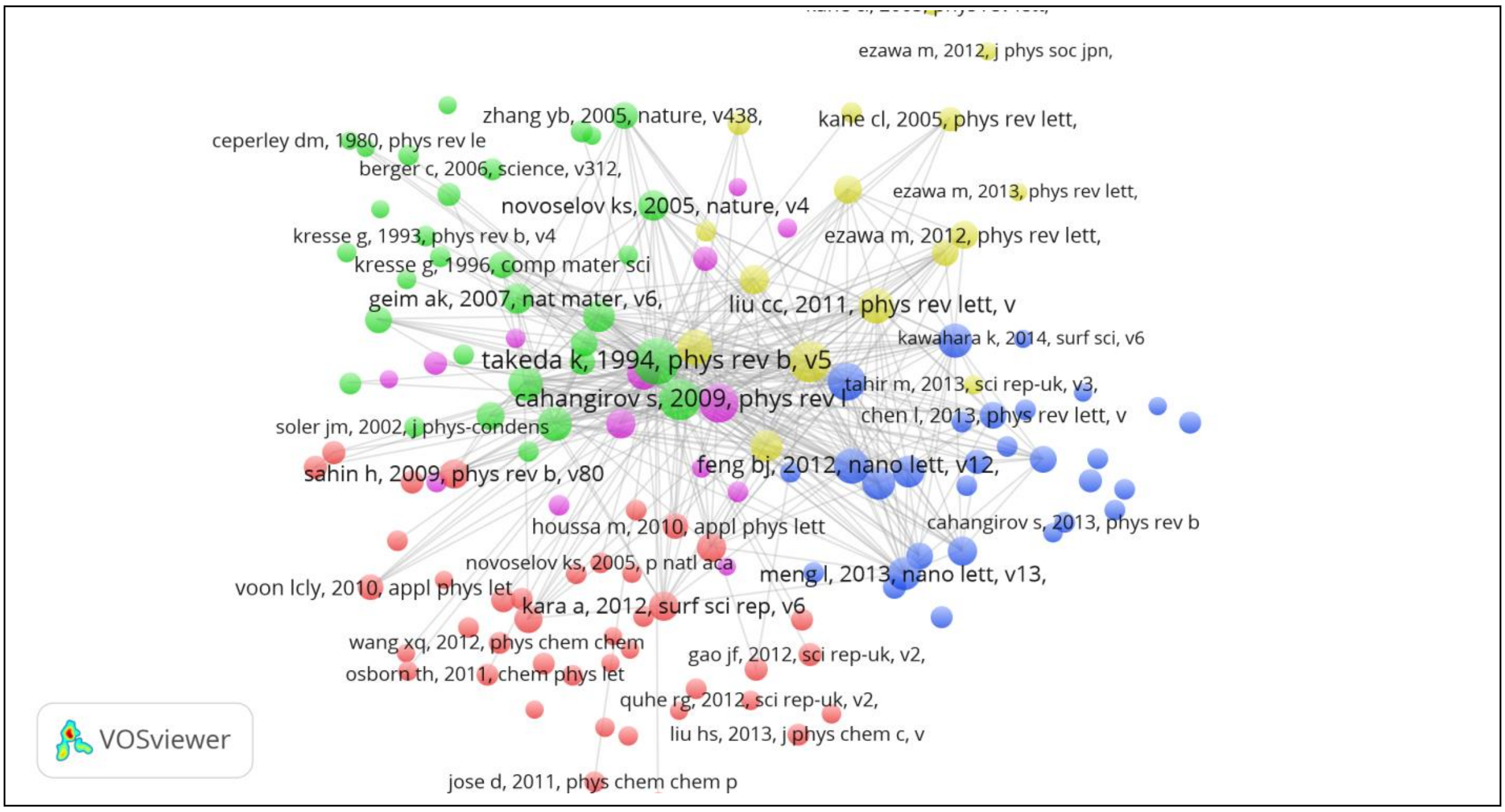

*Figure 20: Co-citation map of the citing papers of the Takeda & Shiraishi SB with links indicating co-citation relations.*

## Concluding Remarks

In this paper we investigated important properties of Sleeping Beauties, particularly to find out to what extent Sleeping Beauties are application-oriented and thus are potential Sleeping Innovations. We found that about half of the SBs are application-oriented. Therefore, it is important to investigate the reasons for, and processes related, to delayed recognition. To this end, we developed a new approach in which the cognitive environment of the SBs is analyzed, based on the mapping of Sleeping Beauties using their citation links and conceptual relations, particularly co-citation mapping. This approach was tested with a blue skies SB and an application-oriented SB. We also found that theoretical SBs are related through many direct and indirect citation links whereas application-oriented SBs are rather isolated. We think that the mapping procedures discussed in this paper are not only important for bibliometric analyses. They also provide researchers with useful, interactive tools to discover both relevant older work as well as new developments, for instance in themes related to Sleeping Beauties that are also Sleeping Innovations.

In order to gain more understanding of the awakening process we will further investigate the characteristics of top-10 SBs. Also we will apply the approaches described in this paper to other fields of science, including an extension our analysis of chemistry and engineering & computer. At the same time, the authors of a number of SBs will be asked to comment on our findings and particularly their views on their 'own awakening'. Other important items are the investigation of as recent as possible SBs, and a further refinement of the mapping techniques to discover why and how a SB deviated from the main research stream in her own direct cognitive environment.

### *Additional data en results for Chemistry and Engineering*

This file can be accessed via http://www.cwts.nl/tvr/ as from July 1st, 2015.

### *Acknowledgements*

The author thanks his colleague Nees-Jan van Eck for developing and writing the Sleeping Beauties algorithm.